\documentclass[%
 aip,
 jap,
 amsmath,amssymb,
 reprint,%
]{revtex4-2}
\usepackage[utf8]{inputenc}
\usepackage{graphicx}
\usepackage{color}
\graphicspath{ {Figures/} } 

\usepackage[caption=false,font=footnotesize,position = top]{subfig}
\usepackage{soul}
\usepackage{chemformula}
\usepackage[english]{babel}
\begin{document}

\preprint{AIP/123-QED}

\title[]{Principles of spintronic THz emitters}

\author{Weipeng Wu}
\affiliation{Department of Physics and Astronomy, University of Delaware, Newark, DE 19716 USA}
\author{Charles Yaw Ameyaw}%
\author{Matthew F. Doty}
 \email{doty@udel.edu}
\affiliation{Department of Materials Science and Engineering, University of Delaware, Newark, DE 19716 USA}
\author{M. Benjamin Jungfleisch}
 \email{mbj@udel.edu}
\affiliation{Department of Physics and Astronomy, University of Delaware, Newark, DE 19716 USA}%
\date{\today}

\begin{abstract}
Significant progress has been made in answering fundamental questions about how and, more importantly, on what time scales interactions between electrons, spins, and phonons occur in solid state materials. These complex interactions are leading to the first real applications of terahertz (THz) spintronics: THz emitters that can compete with traditional THz sources and provide additional functionalities enabled by the spin degree of freedom. This tutorial article is intended to provide the background necessary to understand, use, and improve THz spintronic emitters. A particular focus is the introduction of the physical effects that underlie the operation of spintronic THz emitters. These effects were, for the most part, first discovered through traditional spin-transport and spintronic studies. We therefore begin with a review of the historical background and current theoretical understanding of ultrafast spin physics that has been developed over the past twenty-five years. We then discuss standard experimental techniques for the characterization of spintronic THz emitters and – more broadly – ultrafast magnetic phenomena. We next present the principles and methods of the synthesis and fabrication of various types of spintronic THz emitters. Finally, we review recent developments in this exciting field including the integration of novel material platforms such as topological insulators as well as antiferromagnets and materials with unconventional spin textures. 
\end{abstract}
\keywords{}

\maketitle
\begin{figure}[t]
    \centering
    \includegraphics[width=0.9\linewidth]{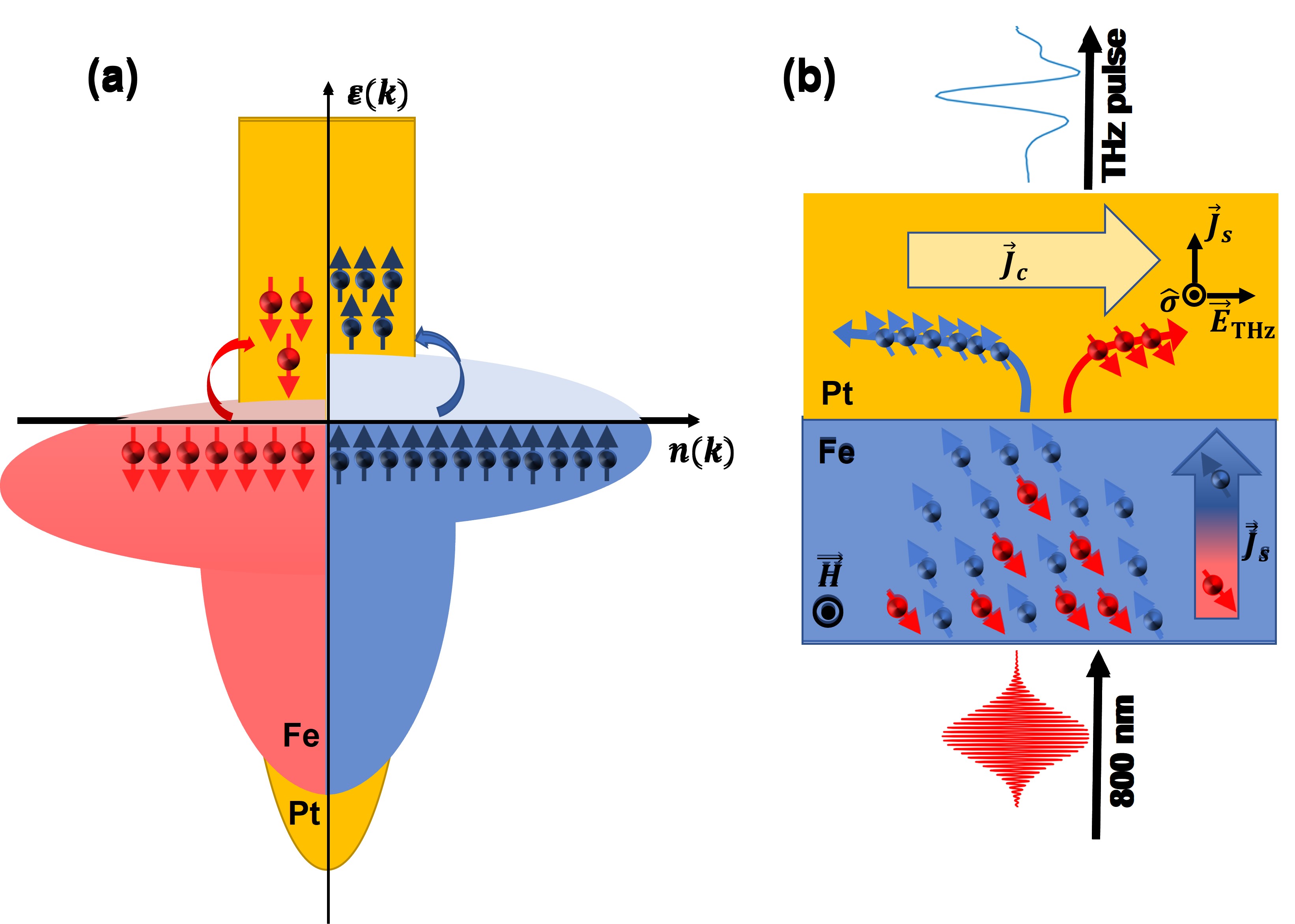}
    \caption{Conceptual overview of THz emission from spintronic devices. (a) Laser excitation of a magnetic material (e.g. Fe) creates more high-energy (mobile) majority spins than minority spins because more majority carriers are available near the Fermi level. (b) Majority and minority spins travel to the interface with a normal metal (e.g. Pt). Majority and minority spins travel in opposite directions along the interface due to the inverse spin Hall effect (and other similar effects). The imbalance in the number of majority versus minority spins results in the net transient charge current that is responsible for THz emission. {The injected spin current propagates normally to the Fe/Pt interface, denoted as ${\vec{J}_s}$, and the direction of spin polarization vector ${\hat{\sigma}}$ is parallel to the external field ${\vec{H}}$ (out of the plane). Owing to the inverse spin Hall effect, ${\vec{J}_s}$ is converted into a charge current ${\vec{J}_\mathrm{c}}$ perpendicular to both ${\vec{J}_\mathrm{s}}$ and $\hat{\sigma}$. This charge current gives rise to a THz transient according to ${\vec{E}_\mathrm{THz}\propto \partial\vec{J}_\mathrm{c}/\partial t}$ as shown in the figure.}}
    \label{fig:THz_conceptual_fig}
\end{figure}

\section{Introduction}
The terahertz (THz) frequency regime is situated between the infrared and the microwave regions of the electromagnetic spectrum, with frequencies spanning from 0.1 THz to 30 THz\cite{miles2001terahertz}. THz frequency radiation has many important applications. For example, many promising materials and molecules for applications in biology and medicine have vibrational rocking and torsion modes that result in optical absorption lines in the THz frequency regime\cite{miles2001terahertz}. These absorption lines serve as a fingerprint enabling detection of the materials through a characteristic THz absorption spectra\cite{miles2001terahertz}. THz radiation is also of great interest for security applications, for example screening for concealed weapons in public spaces such as airports and detecting explosives or life-threatening liquid chemicals in small packages\cite{liu2007terahertz}. These and other applications have motivated significant attention to the THz band of the electromagnetic spectrum from researchers around the world.

One of the most common methods employed in THz research is an absorption spectroscopy technique known as Time Domain Terahertz Spectroscopy (TDTS). A typical TDTS setup consists of a THz source that generates THz pulses and a detector that measures the THz pulse after it has passed through a sample\cite{liu2007terahertz}. We provide a detailed explanation of TDTS in Sect.~\ref{sec:traditional_THz_devices}, but even this simple conceptual overview illustrates an important point: THz sources that are broadband and high-power are an essential component for THz technologies. The two most common methods of THz generation in use today are based on photoconductive antennas (PCA) and optical rectification. We review the operation of these conventional THz sources in Sect.~\ref{sec:traditional_sources}. Here we note that both of these methods take advantage of only the charge and not the spin degree of freedom for electrons\cite{seifert2016efficient}. In recent years, there has been tremendous progress in the field of spintronics and magnetism research that has led to the exploitation of the spin degree of freedom in magnetic materials and composites and to the emergence of new \textit{spintronics}-based THz emitters \cite{Kampfrath_Nat2013,Walowski_JAP2016,Feng:2021ck,Papaioannou_2021}. Conceptually, the operation of a spintronic THz emitter is relatively simple, as depicted in Fig.~\ref{fig:THz_conceptual_fig}. First, excitation of a magnetic material by a near-infrared femtosecond laser pulse generates ultrafast transient spin currents. These ultrafast transient spin currents are converted to ultrafast charge currents at an interface with a proximate material by, for example, the inverse spin-Hall effect\cite{seifert2016efficient}. The ultrafast charge current, in turn, generates THz radiation. Although all THz spintronic emitters operate according to this simple overarching principle, there are many physical effects that can be exploited. The development of increasingly complex materials that control the generation of ultrafast transient spin currents and their conversion to ultrafast transient charge currents thus provides a unique opportunity to engineer the power, spectral width, or pulse shape from THz emitters. We note that there have not yet been any reported spintronic detectors. We therefore focus this article on spintronic terahertz emitters, the physical principles upon which they work, and how the exploitation of electron spin allows one to improve the functionality of THz sources.

This tutorial article is structured as follows: Section~\ref{sec:traditional_THz_devices} reviews traditional THz sources such as photoconductive antennas and optical rectification based on nonlinear crystals, THz detection based on photoconductive antennas and electro-optical sampling, and standard experimental techniques including time-domain terahertz spectroscopy and time-resolved magneto optical Kerr effect measurements. Section~\ref{sec:history} places the development of ultrafast and THz spintronics in a historical context starting from the very first discovery of ultrafast magnetic phenomena in the 1990s. In Sec.~\ref{SpintronicTHzDevices}, we introduce important spintronic effects that were first observed in spin-transport and microwave spectroscopy measurements. Section~\ref{SpintronicTHzDevices} also discusses the synthesis and fabrication of spin-based THz sources and summarizes pioneering works and recent discoveries in the field of THz spintronics. In Sec.~\ref{sec:outlook}, we provide a perspective on ongoing developments, challenges, and opportunities for the future. 

\section{Overview of traditional terahertz generation, detection, and applications}
\label{sec:traditional_THz_devices}
\subsection{Traditional terahertz sources}
\label{sec:traditional_sources}

Historically, the most common source of THz radiation was ``far-infrared sources'' that relied simply on black body radiation. Today, the two most common and convenient methods of THz generation are based on photoconductive antennas (PCAs) and optical rectification. In this section, we summarize the operating principles and the strengths and weaknesses of these two traditional THz sources. Our goal is to explain the current state of THz technology that motivates the interest in spintronic THz sources and to provide benchmarks for the performance characteristics that will make spintronic THz emitters technologically advantageous.

\subsubsection{Photoconducting antennas}\label{Sec:PCA}
\begin{figure}[t]
    \centering
    \includegraphics[width=\linewidth]{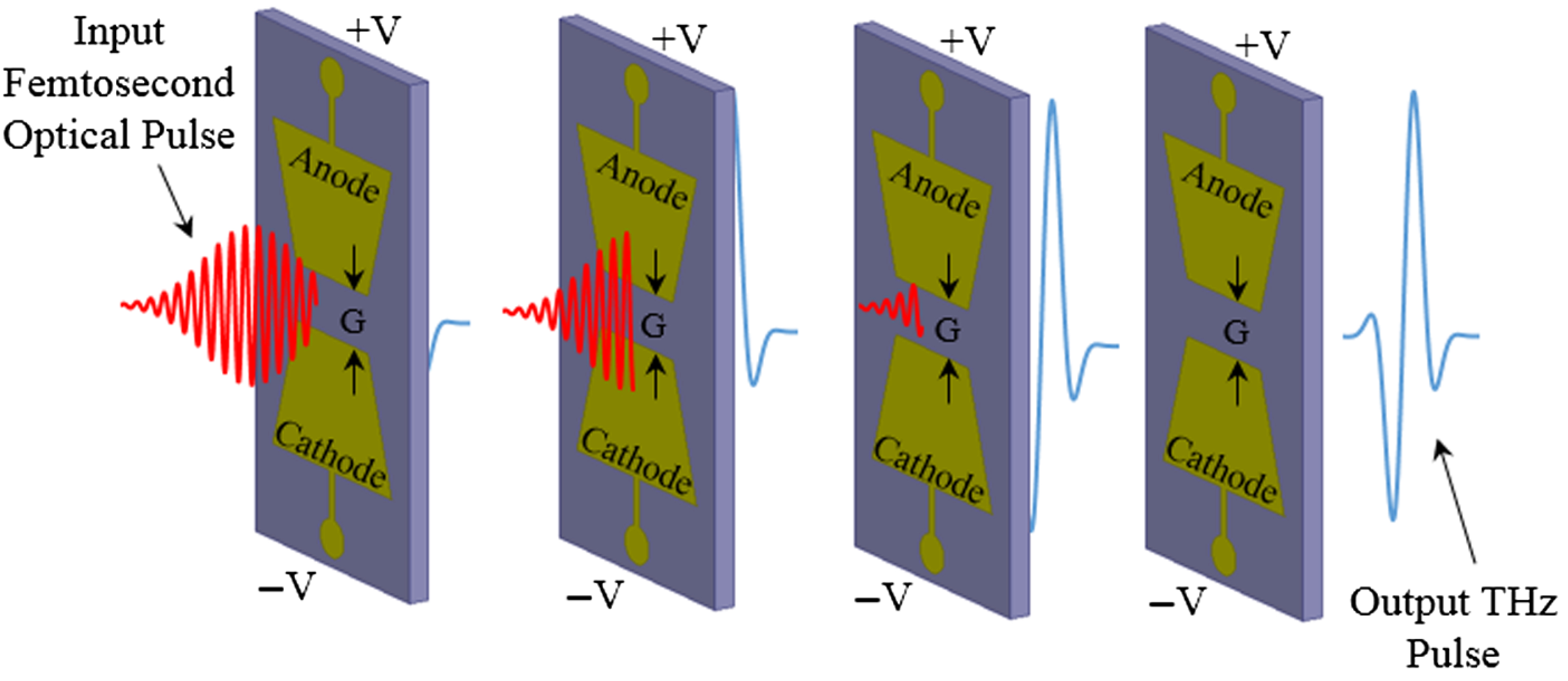}
    \caption{A schematic depiction of THz radiation emitted from a PCA as charges excited by the incident NIR optical pulse generate carriers that accelerate in response to the applied bias voltage. {Reproduced with permission from Optical Engineering \textbf{56}, 010901 (2017). Copyright 2017 Society of Photo-Optical Instrumentation Engineers (SPIE).}}
    \label{fig:PCAemission}
\end{figure}

A photoconductive antenna for terahertz radiation basically consists of a semiconductor thin film of high resistance and two electrical contacts. As shown in Fig.~\ref{fig:PCAemission}, the electrical contacts surround a region of the semiconductor film that is illuminated by an ultrafast optical pulse. The basic operating principle of a PCA, when used as a THz emitter, is that the ultrafast optical pulse generates carriers that accelerate due to the applied bias. The resulting transient charge current generates the THz emission. PCAs can also be used as THz detectors, which we discuss in Sect.~\ref{PCA detectors}. In this section we provide a more detailed description of PCA emitter operation and the impact of the device and operating parameters on the resulting THz emission. 

The semiconductor thin film most commonly used in a PCA is Gallium Arsenide (GaAs), a III-V semiconductor material that is typically epitaxially-grown on a highly-resistive semi-insulating GaAs substrate. The GaAs between the electrodes is illuminated by a pulse of near-infrared (NIR) radiation with temporal width less than 1 picosecond. Because the energy of the laser pulse is larger than the bandgap of the semiconductor material, the photons are absorbed, generating electrons in the conduction band and holes in the valence band. The optically-generated carriers are then accelerated by the electric field created by the biased electric contacts, resulting in the generation of electromagnetic radiation in the THz regime. {While the optical pulse is short (typically $\sim$150 fs) relative to the period of electromagnetic waves at THz frequencies ($\sim$1 ps), generation of a spectrally-broad THz pulse requires not only that the charges accelerate in response to the applied bias but also that the number of carriers decreases rapidly.} To achieve this, the GaAs thin film is typically grown at low temperatures (LT-GaAs) to incorporate a large number of crystal defects that enhance non-radiative recombination of the optically-excited electrons and holes \cite{huang2011terahertz}. From the point of view of a detector, it is similarly important that the optically-generated charge carrier population decays rapidly so that the measured voltage is proportional to the THz electric field at a specific moment in time.

The earliest demonstration of THz radiation generation using photoconductive antennas was conducted in the late 1980s by pioneers David Auston and Daniel Grischkowsky, who used Argon ion-irradiated crystalline silicon epitaxially grown on sapphire \cite{auston1984picosecond}. Since the 1990s, researchers have focused more on III-V materials such as GaAs, InGaAs, and alternating nanoscale multilayers of InGaAs and InAlAs\cite{warren1991subpicosecond}. 
The carriers in GaAs PCAs are typically excited by NIR laser pulses with photon energy larger than the GaAs bandgap (1.42 eV / 870 nm at room temperature). InGaAs-based PCAs are of particular importance for use with fiber-based laser systems. The use of fiber-based lasers can make systems more compact, reliable, and robust, but most fiber-based lasers generate pulses at wavelengths of about 1.55 $\mu$m. InGaAs has a small bandgap energy of 0.8 eV (1.55~$\mu$m) and is thus the preferred material for use in fiber-based systems \cite{wood2010terahertz}. More complex structures that combine layers of multiple materials can have similar or even better performance than bulk InGaAs or bulk GaAs alone \cite{sartorius2008all}. 

Another PCA design parameter that affects overall performance is the antenna design\cite{winnerl2008generation}. By antenna design we mean both the distance between the electrodes (gap size) and the number, shape, and configuration of the electrodes. Stone et al.~performed a systematic characterization of the THz radiation from PCA emitters as a function of antenna design\cite{stone2004electrical}. They found that PCAs with smaller gap sizes ($5-50$ $\mu$m) emitted THz radiation with a larger spectral range (bandwidth) than PCAs with medium or large gap sizes ($0.1-5$ mm). The larger bandwidth is attributed to the faster electric field screening in the PCAs with smaller gap sizes \cite{madeo2010frequency}. Stone et al.~found that PCAs with a fixed 500 $\mu$m gap size emitted THz radiation with a spectral range that was independent of the electrode shape (pointed / bow-tied vs. square or round)\cite{stone2004electrical}. This indicates that the spectral range of a PCA depends primarily on the gap size, substrate properties such as the carrier lifetime, and laser source properties such as pulse width\cite{stone2004electrical}. However, Stone et al. also found that PCA emitters with a pointed (bow-tied) antenna geometry emitted THz radiation with a larger integrated intensity of the FFT spectra, meaning that there was increased intensity at almost all frequency components\cite{stone2004electrical}. This increased intensity results in a ``broader'' usable spectral range of the THz emission.

The most common PCAs in use today are made from LT-GaAs and use bow-tie or parallel stripe antenna designs\cite{cai1997design}. While these PCAs are certainly effective, the potential uses of THz technology described above are motivating interest in THz sources with increasingly higher bandwidth and power. There are four primary limitations to what can be achieved with LT-GaAs PCAs. First, most GaAs-based PCAs have a relative poor optical pump-to-THz emission conversion efficiency\cite{ferguson2002materials}. Second, the THz emission intensity tends to saturate at higher optical pump power\cite{moon2014generation,benicewicz1994scaling}. Third, GaAs-based PCAs rarely emit usable intensities at THz frequencies exceeding 4 THz\cite{DreyhauptAPL2005, KlattOpticsExpress2009, GlobischJAP2017}. Fourth, the bandgap of GaAs restricts these emitters to use with NIR lasers emitting at wavelengths shorter than 870 nm. These limitations largely stem from the band structure and carrier dynamics within the GaAs substrate. There has been substantial progress in the development of new III-V materials such as GaBi$_x$As$_{1-x}$, which reduces the bandgap, and ErAs:GaAs or TbAs:GaAs, in which the rare-earth (e.g.~ErAs) forms nanoinclusions that mediate extremely fast carrier relaxation\cite{Azad2008, Bertulis, Bomberger2016a, Bomberger2016,Bomberger2017, Bomberger2015, Cassels2011, Chen2007, Chen2006a, Nathan2007, OHara2006, Vanderhoef2014}. These materials offer significant opportunities for increased control over band structure and carrier dynamics within the PCA substrate. Spintronic materials offer an entirely different material platform with different and complementary carrier dynamics, as we describe below.

\subsubsection{Optical rectification}
Optical rectification is a non-linear process that occurs when an intense AC electromagnetic field (e.g.~laser pulse) is incident on a non-centrosymmetric crystal \cite{wilke2007nonlinear}. Although the laser pulse applies a sinusoidally-varying electric field, the non-centrosymmetric potental results in asymmetric charge displacement and thus the creation of a dipole. When the exciting laser pulses are temporally short and have a correspondingly large spectral bandwidth, the interaction with the crystal leads to a beating of the polarization in the crystal that generates electromagnetic radiation in the terahertz regime \cite{wilke2007nonlinear}. {The conversion efficiency is defined as the ratio of the emitted THz pulse energy to the input pump pulse energy} \cite{koulouklidis2020observation}. Achieving intense THz output via optical rectification requires matching the optical pump group velocity and the terahertz radiation phase velocity, which is typically achieved by selecting appropriate pump wavelengths\cite{yeh2007generation}. Terahertz emission has been reported in a few centrosymmetric crystals, but only when a strong electric field is applied to break the symmetry \cite{wilke2007nonlinear}. 


The pioneering work on optical rectification was done by Bass et al. in 1962 using a 694 nm continuous wave laser incident on potassium dihydrogen phosphate potassium dideuterium phosphate \cite{bass1962optical}. Zernike and Berman also generated a THz signal with a bandwidth of about 3 THz using low-difference frequency mixing of a near-infrared laser in quartz\cite{zernike1965generation}. {More recently, THz radiation has been generated using other nonlinear crystals such as lithium niobate LiNbO3}\cite{Tian2021} {and zinc telluride (ZnTe), with photon conversion efficiencies or quantum efficiencies up to about 45\%}\cite{yeh2007generation}. {Tian et. al demonstrated the generation of a high-field terahertz pulse train via optical rectification in congruent lithium niobate crystals} \cite{Tian2021}{. The crystal is excited by temporally shaped laser pulses and the resultant THz pulse reaches several hundreds of $\mu$J level. However, the THz pulse trains are narrow-band compared with spintronic emitters.} {THz generation has also been reported in nonlinear organic crystals with output power $2-3$ orders of magnitude greater than that achieved in GaAs-based PCAs}\cite{zhang1992terahertz}. {Such nonlinear THz generation materials can produce higher THz electric field intensities than are typically available from PCAs or spintronic emitters, but they also typically require complex and expensive laser systems to generate intense near infrared pump pulses. Spintronic THz sources are unlikely to be used for generation of high-field THz generation} \cite{Seifert2017} {because of the probability that they will be damaged by the intense pump pulses required.}

ZnTe is the most commonly used electro-optic crystal for terahetrz generation because it has a large second order nonlinear optical susceptibility\cite{liu2004generation}. Spectral ranges up to about 4 THz have been reported using ZnTe crystals\cite{liu2004generation}. While increasing the thickness of the ZnTe crystal allows for higher terahertz radiation amplitude, the increasing thickness makes it impossible to retain the required velocity matching between pump pulse and terahertz.\cite{lee2008principle}. The two major limitations of electro-optic crystals like ZnTe as THz sources stem from this velocity matching requirement: the crystal must be thin and are relatively fragile and a specific pump laser wavelength must be used\cite{lee2008principle}. {Table}~\ref{table:1} {shows a summary of different types of terahertz emitters (including PCA, non-linear crystals and spintronic emitters) and their corresponding useable bandwidth and limitations.}  \newline

\begin{table}[h!]
\centering
\begin{tabular}{ |p{2.3cm}||p{1.8cm}|p{4.1cm}|}
 \hline
 \multicolumn{3}{|c|}{Emitter list} \\
 \hline
 Emitter name & Usable bandwidth (THz) & Limitation(s)\\
 \hline
 LT-GaAs (\textbf{PCA})   & 0.1 - 4\cite{DreyhauptAPL2005, KlattOpticsExpress2009, GlobischJAP2017} & Requires pump laser wavelength of 870 nm or shorter.\\ 
 InGaAs (\textbf{PCA}) & 0.1 -6\cite{klatt2010terahertz} & Requires pump laser wavelength of 1.55 $\mu$m or shorter and engineered sample design.\\
 ZnTe (\textbf{OR}) &0.1 - 4\cite{liu2004generation} & The crystal must be thin and is therefore fragile. A high-intensity with specific pump laser wavelength is required. Expensive emitter. \\
 LiNbO$_{3}$\textbf{(OR)} & 0.1 - 4\cite{liu2004generation} & The crystal must be thin and is therefore fragile. A specific pump laser eiavelength is required.\\
 W/CoFeB/Pt \textbf{(STE)} & 0.1 - 30 \cite{Seifert_Nat2016} & Ultrahigh THz field amplitude challenging to achieve; emitter needs to be magnetized.  \\
 \hline
\end{tabular}
\caption{{Overview of different terahertz emitters and corresponding bandwidth as well as their limitations. The abbreviations are: OR - optical rectification, PCA - photoconductive antenna, STE -  spintronic terahertz emitter}}
\label{table:1}
\end{table}

\subsection{Terahertz detection}
There are two common methods of detecting THz radiation: PCAs and electro-optical sampling. Both of these methods implement a {rapid} measurement of the electric field associated with a THz pulse. The full electric field profile of a THz pulse is reconstructed by scanning the sampling time relative to the emission time of the THz pulse using methods described in Sect.~\ref{TDTS}. Finally, the THz spectrum is obtained through a Fourier transform of the temporal electric field profile of the THz pulse. Because there are no reported spintronic THz detection devices, these established THz detection methods are typically used to characterize the THz emitted from a spintronic source. We introduce the design, operation, and limitations of these two THz detector types here. 

{We note that because the THz spectral data is obtained through a Fourier transform, the spectral range that can be detected is limited by the temporal resolution of the time-domain data. The temporal resolution of the time-domain data for PCA detectors is limited by the lifetime of photo-generated carriers, allowing for measurement of spectral bandwidths up to approximately 8 THz. The temporal resolution of the time-domain data for electro-optic sampling is limited by the probe pulse width and the phase matching with the THz pulse, allowing for measurement of spectral bandwidths up to at least 30 THz}\cite{Seifert_Nat2016}.

\subsubsection{Photoconducting antenna THz detectors}\label{PCA detectors}
The process of detecting terahertz radiation using a PCA relies on the same principle as using a PCA as a terahertz source except that there is no external bias applied to the electrodes \cite{fattinger1989terahertz}. Incoming THz radiation is focused on the dipole antenna. In the absence of a NIR gate pulse, there are no free carriers and no current will be measured at the electrodes. When an optical gate pulse generates free carriers, those carriers accelerate due to the electric field of the THz radiation and generate a photocurrent proportional to the instantaneous THz electric field at the antenna\cite{van1990characterization, jepsen1996generation}. By instantaneous we mean that the laser pulse generating the carriers is short in time, as is the lifetime of the optically-generated carriers. Consequently, the voltage measured at the detector is proportional to the THz electric field over the relatively short window of time defined predominantly by the carrier lifetime. By systematically varying the time at which the NIR laser pulse generates carriers, the temporal electric field profile of the incident THz radiation can be mapped up. We discuss this method in more detail in Sect.~\ref{TDTS}.

Using PCA detectors, Kono et al. reported the detection of terahertz radiation up to 20 THz using a 15 fs (ultrashort) light pulse \cite{kono2000detection}. 
They showed that the  overall performance of a PCA detector depends on the width of the incoming laser pulse as well as the carrier lifetime in the substrate material for the PCA detector \cite{kono2000detection}. In short, the factors that improve PCA emitter performance also improve PCA detector performance. 

\subsubsection{Electro-optical sampling}
Electro-optical sampling is a technique based on the linear electro-optic effect (Pockels effect) \cite{wu1995free}. The Pockels effect was named after Friedrich Carl Alwin Pockels who in 1893 observed changes of the refractive index of an optical medium/electro-optic crystal in the presence of an electric field \cite{wu1995free}. When used for THz detection, the presence of an electric field from the THz radiation changes the refractive index of a crystal. A NIR laser pulse passing through the crystal undergoes a polarization rotation in response to this change in refractive index and this polarization rotation can be detected by using a balanced bridge photodiode. The efficiency of electro-optic sampling for THz detection depends on a number of factors including the absorption coefficient of the electro-optic crystal, the velocity mismatch between laser pulse and terahertz beam, and the spatial overlap of the terahertz beam and optical pulse \cite{tsuzuki2014highly}. In the quest to solve the problem of velocity mismatch between terahertz beam and laser pulse, Wu et al. used electro-optic crystals like ZnTe and GaP that have a smaller absorption coefficient \cite{wu1995free}. In 2008, Pradarutti et al. investigated and compared the THz detection response of CdTe, GaAs, GaP, and ZnTe at a sampling wavelength of 1060 nm\cite{pradarutti2008highly}. CdTe showed a strong signal detection for applications below 1 THz, while GaP was more sensitive to a broader spectrum. In relation to the response function of a zinc blende electro-optic crystal, Kampfrath et al. showed that thick electro-optic crystals could compete with thin crystals in terms of sampling broadband terahertz pulses and could also provide a flatter frequency response\cite{Grischkowsky_1999,Kampfrath_2007}. In essence, the uncertainties in the thickness of electro-optic crystals, particularly for zinc blende crystals, is not crucial for the shape of the detector response\cite{Grischkowsky_1999,Kampfrath_2007}.
 
\subsection{Experimental techniques and methods} 
We now summarize two experimental techniques that utilize terahertz generation and/or detection. The first technique, time-domain terahertz spectroscopy (TDTS), is routinely used for THz absorption spectroscopy {of a wide variety of materials}. It provides a clear example of a technique that can benefit from spintronic THz emitters with stronger intensity or larger bandwidth. It is also the technique most commonly used to characterize spintronic emitters. The second technique, time-resolved magneto-optical Kerr effect (TRMOKE) provides an example of an emerging experimental paradigm that could also benefit from spintronic THz sources. Moreover, TRMOKE enables the measurement of spin population density and dynamics, and therefore is an important method for understanding the ultrafast spin physics that underlie the operation of spintronic THz emission. {TR-MOKE is based on optical measurement of changes in net magnetization or spin orientation, and can therefore only be applied to materials that have such properties.}

\subsubsection{Basics of time-domain terahertz spectroscopy (TDTS)}\label{TDTS}
The basic principles of operation and types of THz sources and detectors have been summarized earlier in this section. It is important to note that these THz detectors sample the electric field at a specific point in time defined by the arrival of the NIR pulse that gates the detector. A complete picture of the electric field transient as a function of time (i.e. the THz pulse) is obtained by using an optical delay line to scan the detector gate pulse relative to the laser pulse that generates THz emission. The electric field as a function of time is then Fourier transformed to obtain the THz spectrum. This approach is called time-domain THz spectroscopy (TDTS). In this subsection we introduce the experimental setup that is commonly used to perform TDTS measurements and explain how all the parts of the system work together.

\begin{figure}[t]
    \centering
    \includegraphics[width=\linewidth]{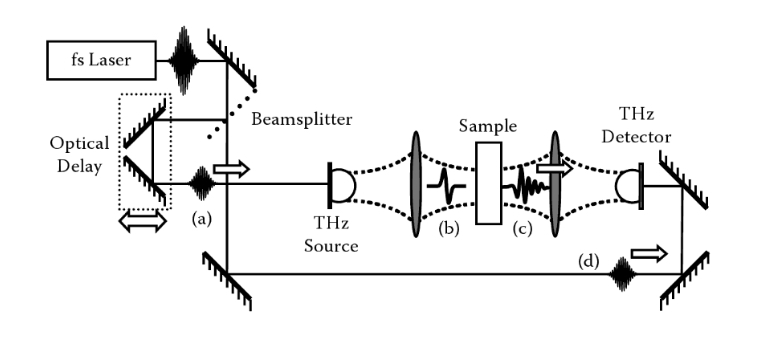}
    \caption{Schematic depiction of TDTS experimental setup as described in the text. {Reproduced with permission from S. L. Dexheimer, Terahertz spectroscopy: principles and applications (CRC press, 2017). Copyright 2017 Taylor \& Francis Group LLC.}}
    \label{fig:TDTSnew}
    \end{figure}
    
A schematic representation of a TDTS setup is shown in  Fig.~\ref{fig:TDTSnew}. At the top left of Fig.~\ref{fig:TDTSnew} is depicted a laser source that generates a laser pulse with femtosecond pulse width. {Although typical pulse widths of the femtosecond pulse laser used are in the range $100-200$ fs,} \cite{dexheimer2017terahertz} {lasers with shorter pulse widths have also been used for THz generation from spintronic emitters} \cite{Seifert_Nat2016,Qiu2020,Weipeng_Wu_JAP2020,Kumar2021}. The laser pulse is split by a beam splitter into a pump beam and probe beam. The pump beam induces THz emission at a source (e.g., a biased PCA) and the probe beam gates the detector (e.g., an unbiased PCA). It is crucial to ensure that the time of arrival of the probe pulse that gates the detector coincides with the time of arrival of the terahertz pulse \cite{dexheimer2017terahertz}. To achieve this, the total optical path length of the two arms, including both the NIR and THz propagation, must be equal. We reiterate that the NIR probe pulse gates the detector and enables a measurement of the THz electric field at that precise moment in time. Thus by routing either the pump or probe beam through an optical delay line, the time delay between THz emission and THz electric field detection can be varied, allowing a measurement of the complete THz electric field in the time domain. No detector has infinitely fast response, and thus the measured signal is actually a convolution of the real THz electric field with the temporal precision of the detector, which is defined, in the case of a PCA antenna, predominantly by the carrier lifetime. The \emph{response function} of a PCA detector is a measurement of the detector's response to an impulse function and is used to deconvolve the actual terahertz field from the measured signal\cite{van1990characterization}.

It is very important to ensure that noise levels within a TDTS system are reduced to a level that ensures accuracy in taking measurements\cite{neu2018tutorial}. The primary source of noise in a TDTS system is laser source noise. When using PCA sources and detectors, for example, an increase in laser fluence would correspond to an increased intensity of THz emission and an increased photocurrent at the detector even if the THz electric field were constant. Stable laser sources are thus important to mitigating noise. Most TDTS experiments utilize lock-in measurement techniques in order to significantly improve signal to noise.

\subsubsection{Time-resolved magneto-optical Kerr effect (TRMOKE)}
The Kerr effect and the Faraday effect are magneto-optical effects that were discovered in the 19th century and have proven to be extremely important as non-destructive probes of the magnetic and spintronic properties of materials. They have been used, for example, to image magnetic domains and spin dynamics in ferromagnetic materials\cite{barman2008benchtop}. Both techniques are based on rotations of the polarization of light when interacting with a material that has a nonzero magnetization or spin projection, and both can be understood as arising from asymmetries in the interaction with the material of the left- and right-circularly polarized components of the polarization. In the Magneto-Optical Kerr Effect (MOKE) the polarization rotation is measured upon reflection from the sample\cite{zhang2009paradigm}. In the magneto-optical Faraday effect, the polarization rotation is measured upon transmission through the sample\cite{barman2008benchtop}. We focus here on MOKE, which is most commonly used for magnetic samples that are, in general, not transparent to the wavelength of light used. 

{While neither MOKE nor TRMOKE are commonly used for THz emission or detection, they are routinely used as a means of characterizing the spin dynamics that underlie the generation of ultrafast transient spin currents and the conversion of such transient spin currents into ultrafast transient charge currents. They are therefore important techniques for understanding the physics that underlay spintronic THz emitters.} To understand MOKE conceptually, consider plane-polarized light that reflects off the surface of a sample. This linearly-polarized light can be decomposed into equal amplitudes of left- and right-circularly polarized light and these left- and right-circularly polarized components will interact differently with magnetization pointed along or perpendicular to the optical propagation direction. The result is that the reflected beam is elliptically polarized. The degree of ellipticity and the Kerr angle (the angle of the major axis of the ellipse relative to the incident plane of polarization) are proportional to the magnetization in the sample\cite{barman2008benchtop}.

There are three important geometries for a typical MOKE system: longitudinal, transverse and polar. These geometries are defined by the relationship between the plane of polarization of the incident light and the magnetization in the sample\cite{barman2008benchtop}. In the longitudinal geometry, the magnetization vector of the sample lies in-plane and is parallel to the plane of polarization of the incident light. In the transverse geometry, the magnetization vector is parallel to the sample plane but perpendicular to the plane of polarization of the incident light. In polar MOKE, the magnetization vector is perpendicular to the sample plane. Vector MOKE methods have been developed to measure the sample magnetization along all three directions by using various combinations of linear and circular polarization in the incident and detected light \cite{Keatley2009, Fan2016, Celik2019}. While the details of these MOKE geometries are not critical to the main point of this article, it is useful to know, in the context of the magnetization and spin dynamics that underlie spintronic THz emission, that it is possible to measure the full three-dimensional dynamics of spins by combining these methods.   

\begin{figure}[t]
    \centering
    \includegraphics[width=\linewidth]{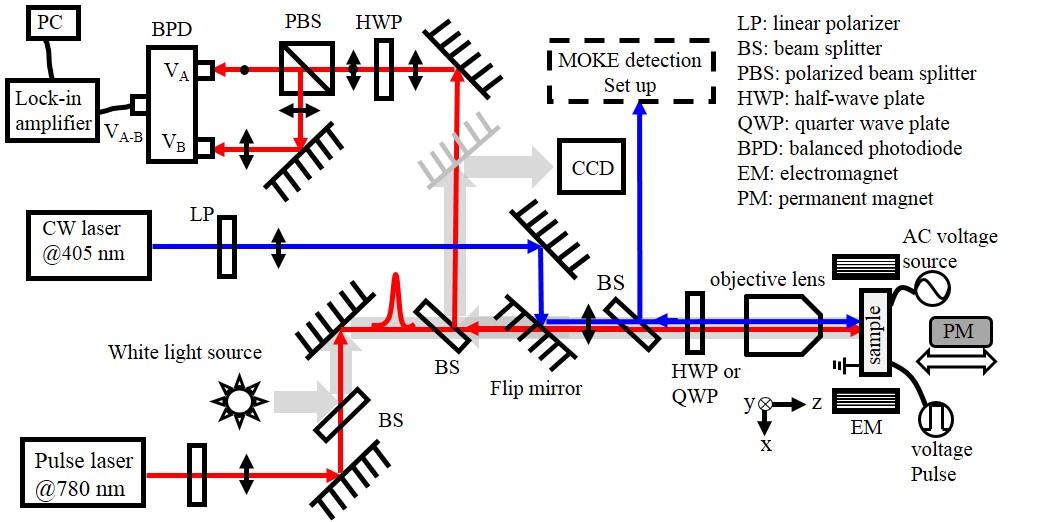}
    \caption{Schematic diagram of a two-color vector MOKE system including TRMOKE capabilities. The red line indicates the beam path for the 780~nm TRMOKE setup, and the blue line represents the 405~nm quasi-static MOKE setup. {Reproduced from Ou et al., Review of Scientific Instruments \textbf{91}, 033701 (2020), with the permission of AIP Publishing.}}
    \label{fig:TRMOKE}
\end{figure}

MOKE measurements were originally developed using continuous wave (CW) lasers in which the MOKE signal is proportional to the average magnetization over the measurement interval. Time-resolved MOKE (TRMOKE) utilizes MOKE detection with a pulsed laser in order to measure the magnetization during a short temporal window, typically of order 150 fs and defined by the temporal width of the optical pulse employed. By scanning the MOKE measurement time relative to any driving function that induces a change in the magnetization, TRMOKE can be used to map the temporal evolution of the magnetization, including both oscillations and damping to new equilibrium values\cite{zhou2020investigation,barman2008benchtop,ou2020development}. A schematic representation of a TRMOKE system is shown in Fig.~\ref{fig:TRMOKE}. The laser systems and optics used for a typical TRMOKE system are quite similar to those used in TDTS. One additional requirement for TRMOKE is that the driving function that induces magnetization changes must be synchronized to the laser measurement pulses\cite{ou2020development}. Moreover, this driving function must have a sufficiently sharp rising edge that all magnetization changes begin at a well-defined point in time relative to the measurement pulse\cite{ou2020development}. A common driving function is a step function in electrical current that drives magnetization re-orientation through spin-orbit torques\cite{ou2020development}. Finally, we note that achieving good signal to noise in TRMOKE measurements typically requires the use of lock-in measurement techniques sampling thousands to millions of repetitions of identical pump-probe conditions. Consequently, system stability and rapid return to equilibrium conditions within the sample are essential. 

{A number of TRMOKE-based pump-probe techniques have been used to study the spin dynamics of magnetic materials and heterostructures. For example, the step function in current described above can be replaced by an optical pulse that induces spin dynamics}. See, for example, the discussion in Sect.~\ref{sec:ultrafast_spintronics} associated with Fig.~\ref{fig:demag2}{, which illustrates how TRMOKE can measure magnetization precession and damping. There are also efforts underway to use THz radiation to induce spin dynamics that can then be measured using MOKE. For example, THz photons incident on a metal grating deposited on a topological insulator can generate propagating Dirac plasmon polaritons}\cite{Wang2020f} {whose dymanics could be measured by a THz pump - MOKE probe system. Such systems would benefit from THz spintronic emitters that can provide higher THz intensities or larger bandwidths.}

\section{The physics of spintronic THz emission}
\label{sec:history}
\subsection{Conceptual overview}
All spintronic THz emitters are based on a sequence of physical processes. First, a temporally-short laser pulse, typically in the NIR, generates a non-equilibrium spin population. Second, the formation of this non-equilibrium spin population results in a transient spin current. Third, the transient spin current encounters an interface where it is converted into a transient charge current. Finally, this transient charge current results in the emission of THz frequency radiation. Although all spintronic THz emitters follow this general sequence, there are a variety of physical effects that can contribute to the precise way in which, for example, the transient spin current is generated. Similarly, there are a variety of physical mechanisms by which the transient spin current can be converted to a transient charge current. Understanding the ways in which materials and interface engineering can be used to control the spin current generation and spin current-to-charge current conversion requires a strong foundation in the underlying physical processes. 

An excellent and thorough summary of THz spintronics and ultrafast magnetism was recently published by Walowski and M{\"u}nzenberg\cite{Walowski_JAP2016}. The purpose of this section is to introduce the physical processes that underpin the generation of ultrafast spin currents and their conversion into transient charge currents, which are the essential steps in spintronic THz emission. In Sect.~\ref{sec:ultrafast_history} we summarize the early ultrafast studies of magnetic materials. We focus on a process known as ultrafast demagnetization, which was understood as the loss of spin angular momentum through transport of spins. In modern language, we would say that this ultrafast demagnetization is a result of a transient spin current, and these transient spin currents are precisely what we want to generate and then convert into transient charge current in order to generate THz radiation. In Sect.~\ref{sec:ultrafast_spintronics} we introduce the concepts and models for the physical processes that underlie the generation of transient spin currents. In Sect.~\ref{SpinToCharge} we discuss the physical process by which this transient spin current can be converted into transient charge current and result in THz radiation. In Sections \ref{SpintronicTHzDevices} and \ref{sec:outlook} we will discuss how materials and heterostructures can be selected and combined to control these physical processes and engineer improved THz emission. The detailed descriptions of the underlying physics in this section are intended to provide the foundation necessary to understand the wide range of materials and interface engineering that can be used to control the spin current and spin-to-charge current conversion in such devices. 

\subsection{The birth of terahertz spintronics: a short summary of ultrafast processes in magnetic materials}\label{sec:ultrafast_history}

In the following we give a brief historical overview of the field of ultrafast demagnetization processes and then describe ultrafast spin transport.

\subsubsection{Ultrafast demagnetization and the three-temperature model}\label{Demag}
Beaurepaire, Bigot, and coauthors were among the first to explore magnetic materials using sub-ps optical spectroscopy methods.\cite{Beaurepaire1996} In this section we first use this work by Beaurepaire as an example to illustrate the historical study of ultrafast demagnetization and the model that was used to understand it. Our discussion of this example also serves to introduce several of the key physical concepts about how magnetic materials respond to ultrafast laser pulses.

\begin{figure}[t]
    \centering
    \includegraphics[width=0.8\linewidth]{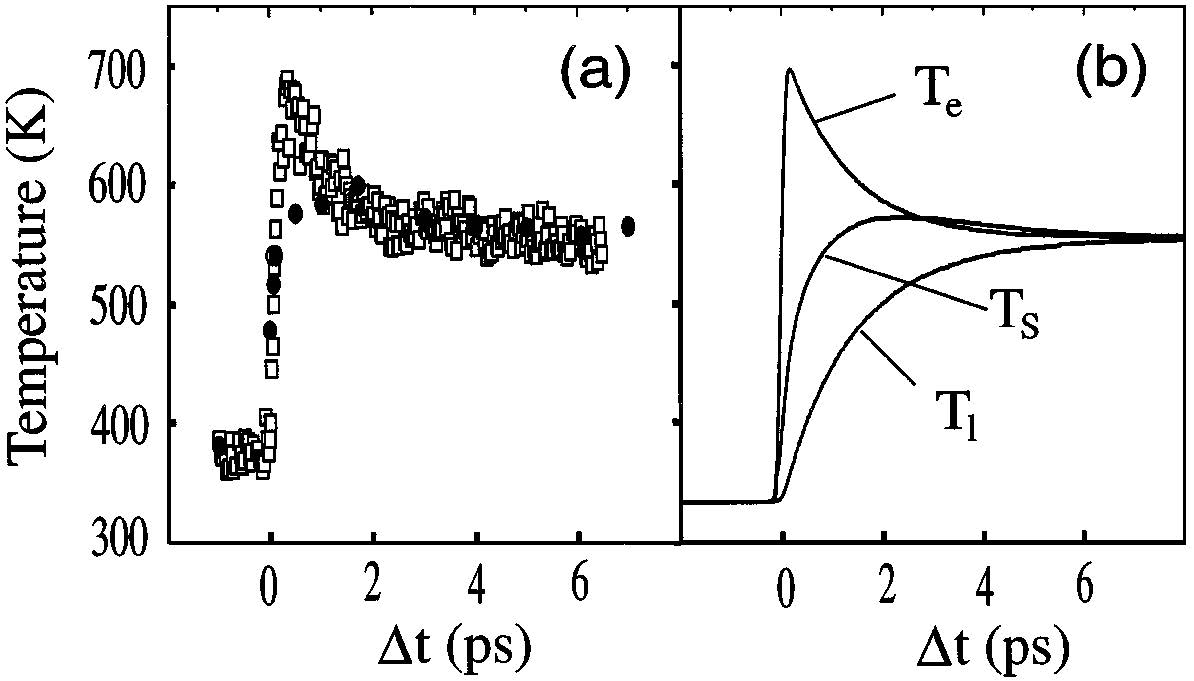}
    \caption{(a) Measured net temperature explained by Beaurepaire et al.~as arising from (b) time-dependent electron ($T_e$), spin ($T_s$) and lattice ($T_l$) temperatures. Reproduced with permission from {Phys. Rev. Lett. \textbf{76}, 4250 (1996). Copyright 1996 American Physical Society.}}
    \label{fig:ThreeTemp}
\end{figure}

Beaurepaire and coauthors excited a thin Ni film with a 60 fs optical pulse at 620 nm and were able to separate the changes to the electron and spin populations by applying complementary magneto-optical and all-optical methods. First, they performed TRMOKE measurements in which they swept the magnetic field intensity and direction to create a Kerr hysteresis loop for each time delay. From these hysteresis loops they extracted the remanent magnetization as a function of time, which allowed them to isolate the transient changes to the spin temperature. Second, they performed transient absorption measurements that were affected only by the electron temperature. From this data, reproduced in Fig~\ref{fig:ThreeTemp}(a), they developed the three-temperature model for interactions between electrons, spins, and phonons. 

The three-temperature model describes the separate evolution of the electron, spin, and lattice temperatures in response to an ultrafast optical pulse. To understand the three-temperature model, we start with the Fermi-Dirac distribution that describes the number of electrons in a material as a function of energy. When a sample is at absolute zero, the Fermi-Dirac distribution is a step function because all electrons have relaxed and filled every allowed state up to the Fermi level. As the sample temperature increase, the distribution spreads out because an increasing number of electrons have energy above the Fermi level. In equilibrium conditions, the electron temperature that creates this distribution is the same as the lattice temperature, which is defined by the amount of energy contained in the vibrations of the lattice (i.e. phonons). When an ultrafast optical pulse is absorbed by a material, this equilibrium is disturbed. Some number of electrons is excited to higher energy levels, increasing the total amount of energy in the electron gas. The electron temperature, $T_e$, is the temperature that corresponds to the Fermi-Dirac distribution of electron energies. Absorption of an ultrafast optical pulse by a material thus promotes a subset of the electron population to higher energy states, creating a non-thermal distribution \cite{Fann.PhysRevB.46.13592.1992, Sun.PhysRevB.50.15337.1994, Suarez.PhysRevLett.75.4536.1995}. Electron-electron interactions rapidly exchange energy among the electrons, creating a thermal distribution at a new (higher) $T_e$. This thermalization of the electron population typically occurs on timescales of a few hundred fs. This rapid increase in $T_e$ is shown in Fig~\ref{fig:ThreeTemp}(b). Over time, interactions between the electrons and the lattice transfer some of this electron energy into phonons, increasing the lattice temperature $T_l$. As can be seen in Fig~\ref{fig:ThreeTemp}(b), the lattice and electron temperatures will come to equilibrium on timescales of order 10 ps.

The spin temperature $T_s$ is a fictitious temperature that is used to describe the relative number of majority and minority spins. As depicted in Fig.~\ref{fig:THz_conceptual_fig}(a), there are more majority spins than minority spins in any magnetic material at equilibrium. As temperature increases, the thermal energy exceeds the energy of the exchange interactions that lead to spin alignment and thus the spins become increasingly equally distributed among the majority and minority orientations. $T_s$ is defined as the temperature that would result in the observed relative number of majority and minority spins. Because optical excitations in metals are spin conserving, the initial change in the energy levels of the electrons, which leads to the changes in $T_e$, does not result in a change in the net spin polarization, i.e. the relative number of majority and minority spins. This process is also depicted in Fig.~\ref{fig:THz_conceptual_fig}(a). After optical excitation, scattering leads to the homogenization of the majority and minority spin populations described by the increasing $T_s$ observed in Fig~\ref{fig:ThreeTemp}(b)\cite{Beaurepaire1996}.

{While the phenomenological description of the three-temperature model successfully captured the main experimental findings, there has been steady progress over the past 25 years in developing more realistic models capable of distinguishing the microscopic processes involved \cite{Schellekens}. For example, in the atomistic approach \cite{Kazantseva,Skubic}, where exchange interaction mediates the ferromagnetic coupling between spins, heating of the electron system by an ultrafast laser pulse results in a random reorientation of individual spins. On a macroscopic level, this leads to a reduction of the magnetic order and thus the macroscopic magnetization vector. The Landau-Lifshitz-Bloch description is similar to the atomistic approach. Here, the thermal response is modelled using the Landau-Lifshitz equation with stochastic behavior of individual spins via a mean-field approximation of the exchange interaction \cite{Atxitia,Atxitia2}. Finally, there has been discussion of a Stoner-like bandstructure approach in which the spin-flip events occur due to scattering with electrons, magnons, and phonons leading to a reduction of the magnitude of the magnetic moments \cite{Carva,Essert,Krauss}.}

\begin{figure}[t]
    \centering
    \includegraphics[width=\linewidth]{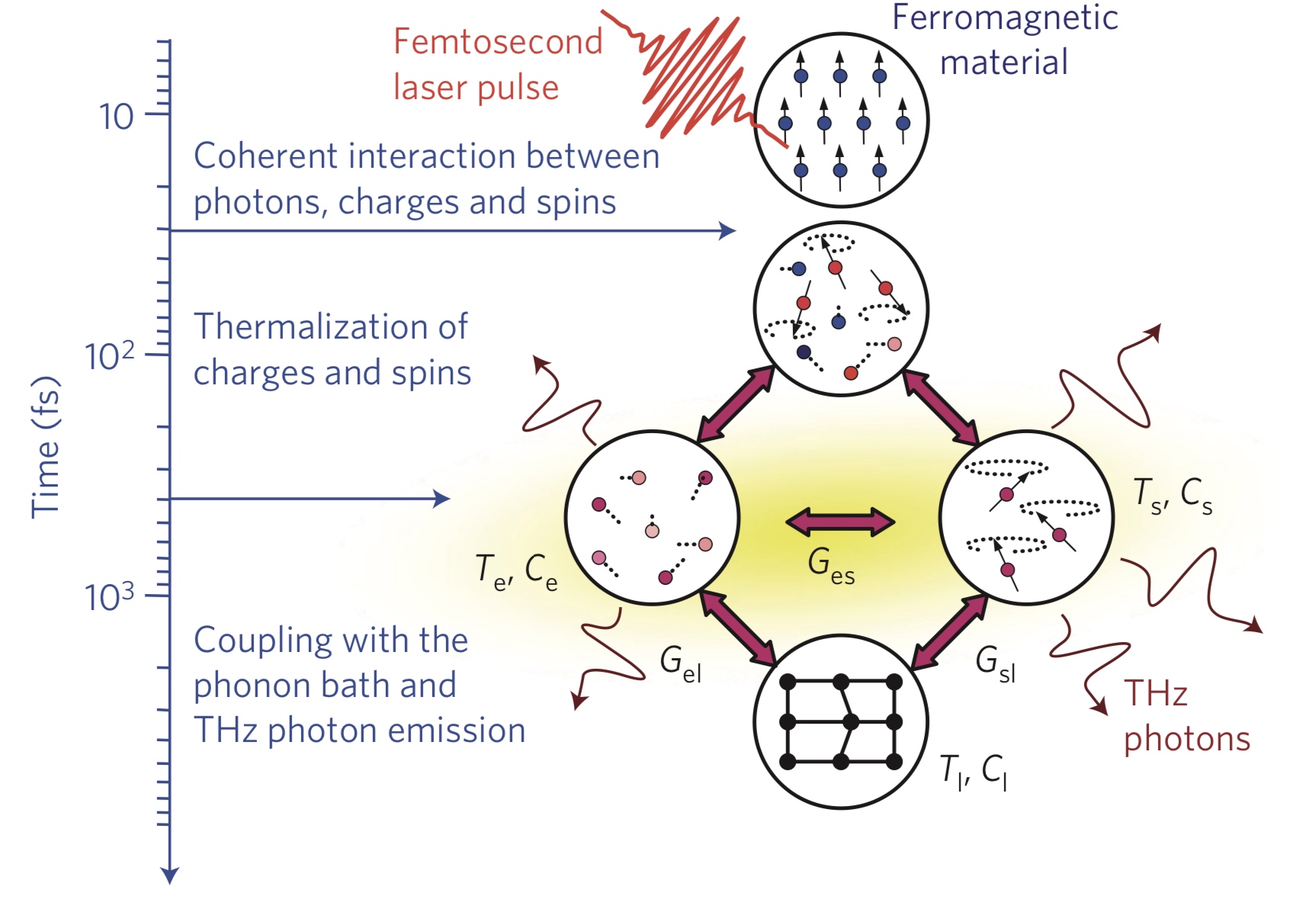}
    \caption{Charges and spins of a ferromagnet are excited by a femtosecond laser pulse that induces a demagnetization after a coherent interaction time. This is followed by a thermalization process, in which charges and spins interact together with the lattice. This is described by coupled baths with temperatures $T_\mathrm{e}$, $T_\mathrm{s}$ and $T_\mathrm{l}$. The coupling constants are denoted $G_\mathrm{es}$, $G_\mathrm{el}$ and $G_\mathrm{sl}$ with $e$ -- electrons, $s$ -- spins, $l$ -- lattice. The demagnetization process is accompanied by emission of terahertz radiation. {Reproduced with permission from Nature Physics \textbf{5}, 515-520 (2009). Copyright 2009 Springer Nature.}}
    \label{fig:timescales}
\end{figure}
More recent work by Beaurepaire, Bigot, and coauthors has established that there are coherent and nonlinear interactions between the electromagnetic field of the exciting laser pulse and the magnetization (spins) within magnetic materials\cite{Bigot2009}. This coherent interaction between the electromagnetic field and the magnetization can be distinguished from the polarization free decay that results from loss of electronic (charge) coherence. As summarized in Fig.~\ref{fig:timescales}, these coherent interactions occur within the duration of the optical pulse ($\sim50$ ps) and are then followed by spin population and thermalization dynamics that can be characterized by the electronic ($T_e$), spin ($T_s$), and lattice ($T_l$) temperatures. As also depicted in Fig.~\ref{fig:timescales}, THz photons can be emitted during the demagnetization process. These results show that the idea that spin angular momentum is lost can be reconciled with the conservation of angular momentum when coherent interactions between the electromagnetic field and the magnetization are included. Moreover, they suggest an opportunity to use tailored light pulses to precisely control magnetization dynamics, which could be used to shape THz pulses.

\subsubsection{Ultrafast spin transport}\label{history_SpinTransport}
The excitation of a magnetic sample by an ultrafast optical pulse, as described above, leads to an increase in the average electron (spin) velocity. There are many different methods for calculating or simulating the resulting electronic and spin transport, but most analyses start from Boltzmann transport\cite{Battiato_PRL2010,Battiato_PRB2012,Nenno_2019,Lu_PRB2020}. The key conceptual idea of Boltzmann transport relevant to spintronic THz emission is that the high density of excited (high momentum) electrons resulting from the optical excitation will result in a flow of electrons (spins) into unexcited regions (e.g. the interface with a normal metal) that have a lower density of excited electrons. We next distinguish between the ballistic and diffusive transport regimes. Ballistic transport describes electrons travelling according to their initial momentum, while diffusive transport is characterized by a strong scattering rate \cite{Walowski_JAP2016}. Right after the laser excitation, electron transport can be described in the ballistic limit. The transport characteristics then gradually change and approach the diffusive regime \cite{Lu_PRB2020} with a time-constant determined by the scattering rate. Superdiffusive spin transport \cite{Battiato_PRL2010,Battiato_PRB2012,Lu_PRB2020} is observed in the transition between the ballistic and diffusive transport regimes: most electrons scatter and create secondary electrons, while some electrons propagate ballistically \cite{Walowski_JAP2016}. The recent work of Nenno and coauthors provides a nice example of how simulations of the hot-carrier dynamics following laser excitation of bilayers containing a ferromagnetic metal (FM) and a normal metal (NM) can be used to predict and understand the resulting THz emission\cite{Nenno_2019}.


\subsection{Spin-polarized currents and spin waves}\label{sec:ultrafast_spintronics}
Spintronics, which refers to spin-electronics, is an emerging field that utilizes the spin degree of freedom to advance traditional electronic concepts. 
While conventional electronic devices are approaching the limits of miniaturization due an increased generation of waste heat, spintronic devices may be able to circumvent these drawbacks. They offer additional functionalities such as higher operational speed and lower power consumption. 

In general, we can distinguish two types of spin currents: spin-polarized electric currents and spin waves (or their fundamental quanta - magnons).  
Spin-polarized electron current is an electric current with an unequal amount of spin-up and spin-down electrons. In contrast, a spin wave is a propagating perturbation in a magnetically ordered material that can be used to transfer spin angular momentum. While spin waves can be used in both ferromagnetic metals and insulators, which is an important advantage when it comes to power consumption, spin-polarized electron currents in metals can be created and controlled more easily. In the next two sections, we will introduce the basics of the two types of spin currents in more detail. 

\begin{figure}[t]
    \centering
    \includegraphics[width=0.9\linewidth]{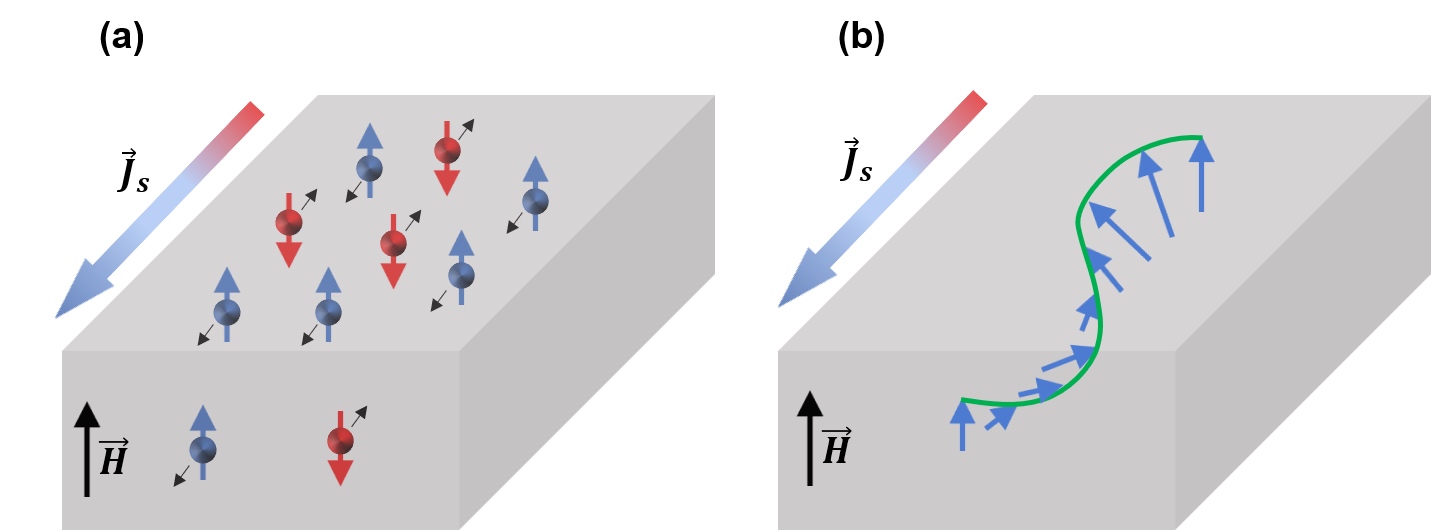}    \caption{Illustration of (a) spin-polarized current and (b) spin waves. The spin polarized current is the flow of electrons with non-zero net spin polarization in a conductor. Spin waves are collective precessional motion of spin angular momentum.}
    \label{fig:my_label}
\end{figure}

\textbf{Spin-polarized electron currents}
J.~C.~Slonczewski \cite{slonczewski_current-driven_1996} and L. Berger \cite{berger_emission_1996} independently proposed in 1995 a new mechanism to control the magnetization of a magnetic material by a spin-polarized current now known as the spin-transfer torque effect. In a ferromagnet/nonmagnetic/ferromagnet trilayer structure, an unpolarized electric current becomes spin-polarized after passing through the ferromagnetic layer. In this process, exchange interaction tends to align the spins of the incoming electrons parallel or antiparallel to the local magnetization. Since there is an imbalance between majority and minority spin electrons in a ferromagnetic metal near the Fermi level, the electrons become spin polarized and hence a spin-polarized current is created \cite{dyakonov_spin-orbit_2017}.
This spin-polarized current traverses the nonmagnetic layer, which is used to avoid exchange interaction between the two magnetic layers. After passing through this layer, the spin current is injected into the second magnetic layer where it exerts a torque on the magnetization. This torque can lead to the onset of a steady state precession or result in switching of the magnetization. This can be considered the inverse of the ``spin filtering'' process occurring in the first magnetic layer. 


Two other possible ways to efficiently create spin-polarized electron currents that also work in insulating magnets are the inverse spin-galvanic effect \cite{New-photogalvanic-effect-in-gyrotropic-crystals, Spin-polarization-of-conduction-electrons, Crowell_2019} and the spin-Hall effect \cite{Dyakonov1971}. As described in more detail in the next section, the inverse spin-galvanic effect is the result of asymmetric spin-flip scattering of electrons in gyrotropic materials due to a broken inversion symmetry in the system \cite{ganichev_spin_2019,thomas2007handbook}, which leads to a homogeneous spin density throughout the sample. The spin-Hall effect occurs in systems where the inversion symmetry is conserved and results in a spin current perpendicular to both the electric charge current and the spin-polarization vector. 

\textbf{Spin waves} Apart from spin angular momentum carried by spin-polarized currents, spin angular momentum can also be carried by spin waves in magnetic media, both metallic and insulating. A spin wave is the propagation of the collective excitations of the spin lattice \cite{Magnetization-Oscillations-andWaves-Gurevich1996MagnetizationOA,spin-waves} in a magnetically ordered material. 
The quanta of spin waves are called magnons \cite{PhysRev.58.1098, PhysRev.102.1217}. Compared with spin currents that rely on the movement of electrons in a conductor, magnonic spin currents hold the promise of being energetically more efficient due to the absence of electron scattering in magnetic insulators and therefore a reduced Joule heating in comparison to conductors. 
\begin{figure}[t]
    \centering
    \includegraphics[width=0.9\linewidth]{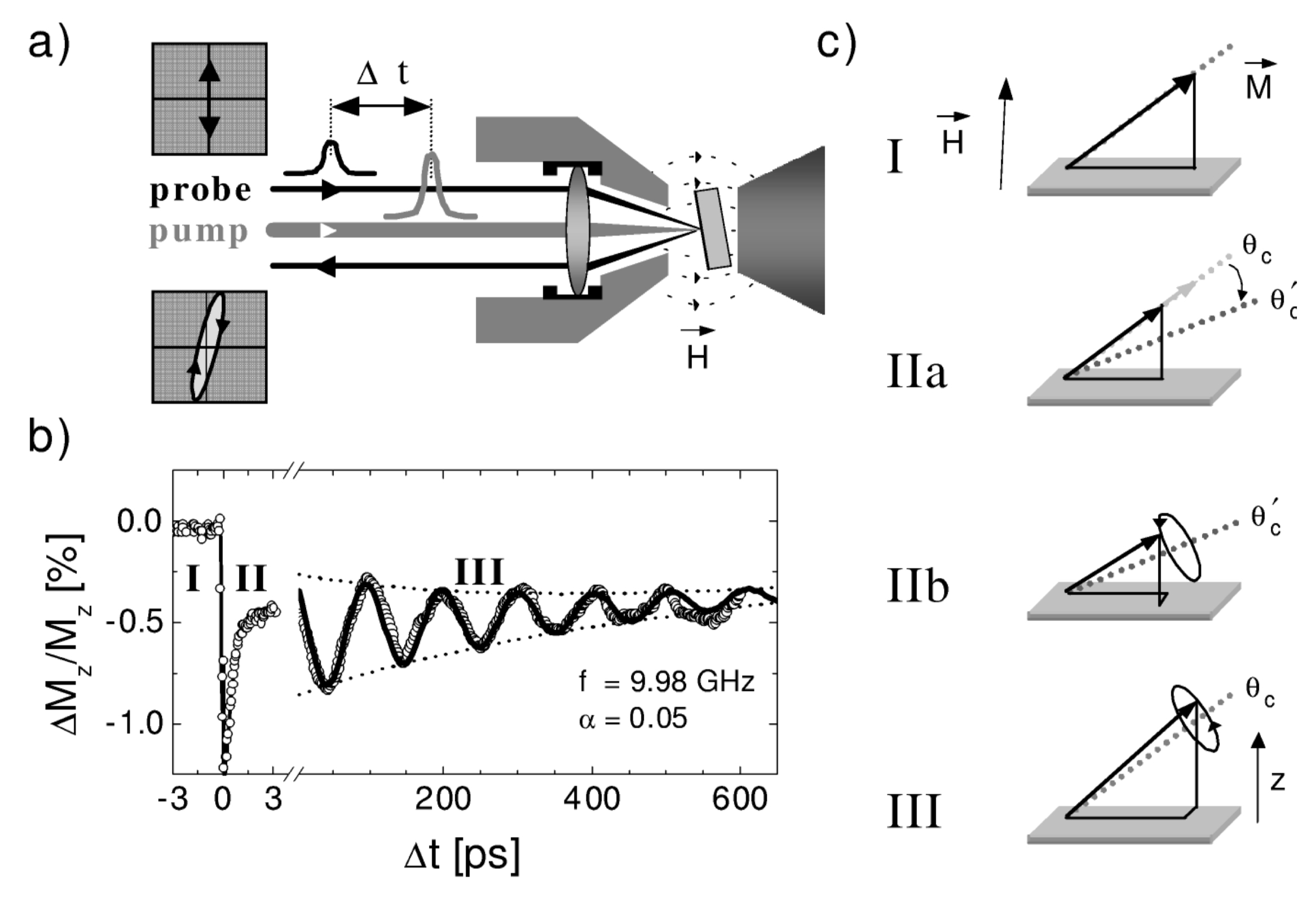}
    \caption{(a) Laser pump pulse excitation of a Ni film results in spin dynamics that are measured with a time-delayed optical probe pulse. (b) The observed spin dynamics can be explained by (c) the initial displacement of the magnetization from equilibrium followed by precession and damping. 
    {Reproduced with permission from Phys. Rev. Lett. \textbf{88}, 227201 (2002). Copyright 1998 American Physical Society.}}
    \label{fig:demag2}
\end{figure}
In the classical limit, the fundamental equation of motion of the precessing magnetization $\vec{M}$ in an effective magnetic field ${\vec{H}_{\rm{eff}}}$ is described by the {Landau–Lifshitz–Gilbert (LLG)} equation \cite{GilbertIEEETM2004}:
\begin{equation}
    \frac{d \vec{M}}{d t}=-\left|\gamma\right|\vec{M}\times\vec{H}_{\mathrm{eff}} + \frac{\alpha_\mathrm{G}}{M_\mathrm{S}}\vec{M}\times\frac{d \vec{M}}{d t}
\label{eq:LLG}
\end{equation}
where $M_\mathrm{S}$ is the saturation magnetization, $\gamma$ is the gyromagnetic ratio, and $\alpha _\mathrm{G}$ is the Gilbert damping constant. The first torque term on the right describes the precession caused by an effective field including external field, anisotropy fields, demagnetizing fields, etc., while the second term represents the damping of the magnetization precession \cite{GilbertIEEETM2004}. The small angle solution to the LLG is known as the Kittel equation. Figure~\ref{fig:demag2}(b) shows a nice example of how the Kittel equation can be used to fit and understand data demonstrating magnetization precession and damping in response to excitation of a Ni film by an ultrafast optical pulse \cite{VanKampen2002}.  

\begin{figure}
    \centering
    \includegraphics[width=0.5\linewidth]{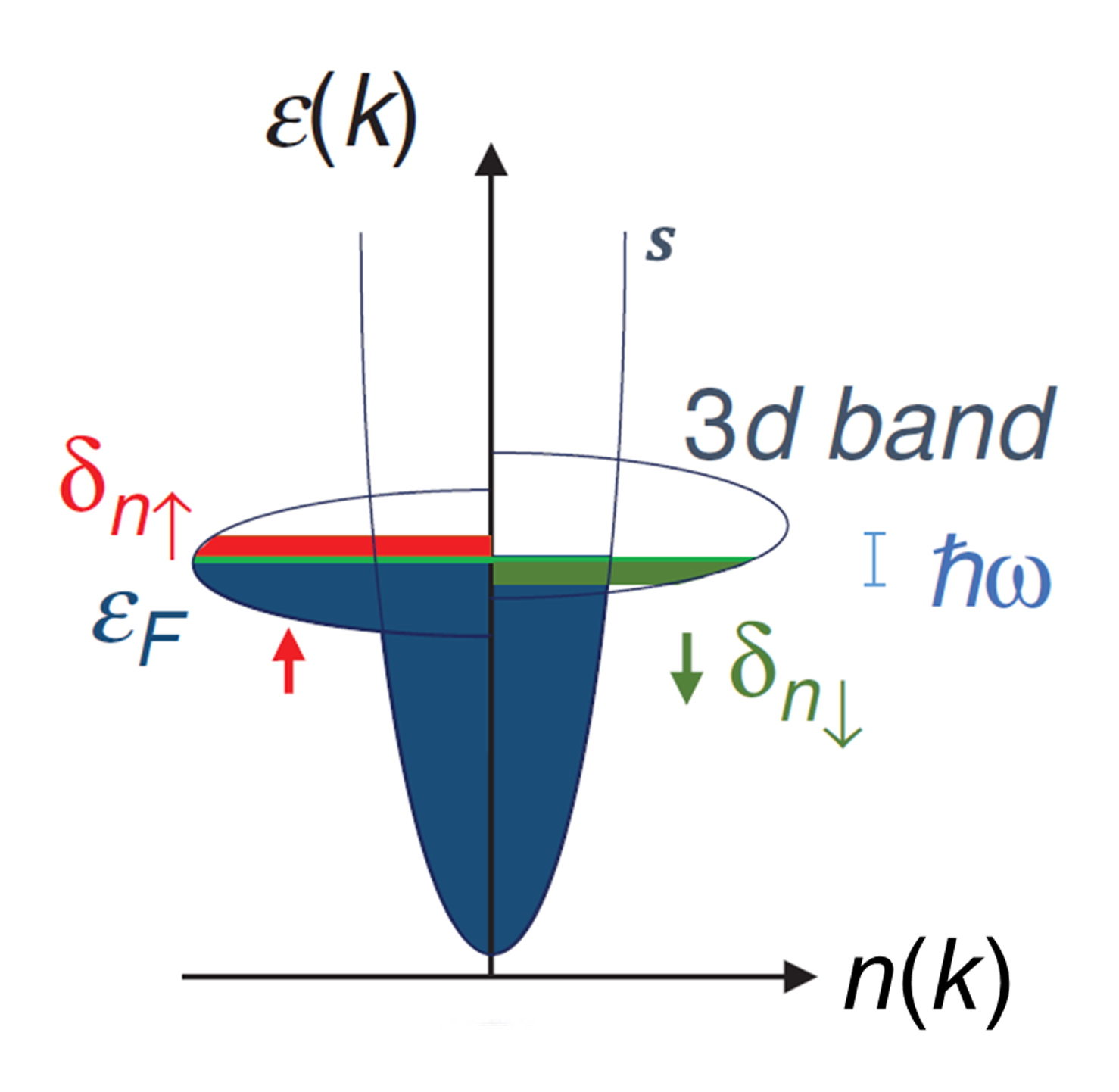}
    \caption{Sketch of the spin pumping effect. The electrons are excited in the GHz frequency (excitation energy $\hbar \omega$) regime near the Fermi level $\epsilon_\mathrm{F}$, leading to the out-of-equilibrium spin-density and spin-current. {Reproduced from T. H. Dang et al., Applied Physics Reviews \textbf{7}, 041409 (2020), with the permission of AIP Publishing.}}
    \label{fig:spin_pumping}
\end{figure}

We note that the precessional motion of the magnetization, such as spin waves, can be converted into spin-polarized currents by the spin-pumping effect: this effect describes the injection of a spin-polarized electron current in a nonmagnetic metal as a result of the magnetization precession in an adjacent magnetic material, either metallic or insulating \cite{Spin_pumping_Tserkovnyak}. The out-of-equilibrium spin density and spin current are generated due to the high-frequency spin dynamics in an energy range close to the Fermi level\cite{spin_pumping_dang}, as illustrated in Fig.~\ref{fig:spin_pumping}. This non-equilibrium spin current can be detected by, for instance, the inverse spin Hall effect, which is discussed in more detail in Sec.~\ref{SpinToCharge}. The generation of a spin current due to an incoherent precession is known as the spin Seebeck effect, which we do not discuss henceforth and refer the reader to the literature \cite{Uchida_2008,Uchida_Nat_Mat_2010,Uchida_APL_2010}. 

{Finally, we note that there has been some very recent efforts in using THz emission from magnetic samples to probe the underlying spin currents. For instance, Zhang et al. showed that it is possible to reconstruct the magnetization dynamics from the THz emission of a ferromagnet, i.e., the demonstration of ultrafast THz magnetometry} \cite{Zhang_NatComm_2020}{, while Qiu et al. demonstrated the direct connection between the angular-dependent THz waveforms and three-fold rotational symmetry of an antiferromagnet, i.e., that it is possible to infer symmetry properties of the system from the THz emission characteristics} \cite{Qiu2020}.

\subsection{Introduction to spin-transport and spin-orbit phenomena}\label{SpinToCharge}
In this section, we give a brief overview of spintronic effects important for the understanding of THz spintronic emitters. For a more detailed review and introduction to those phenomena, we refer the reader to the literature cited in each subsection. 
\newline

\paragraph{Magnetoresistance effects}

The magnetoresistance effect is the change in the electrical resistance when the magnetization state is changed. While the effect was first observed more than 150 years ago, many modern sensor applications still rely on the detection of small changes in magnetoresistance due to small changes in magnetization. There is a large family of magnetoresistance effects \cite{GMR_Baibich,GMR_Binasch,Ennen2016,TMR_1, Liu2017,Kowalska2019} such as giant magnetoresistance (GMR) and tunnel magnetoresistance (TMR) -- to name only a few. We focus here on the description of the anisotropic magnetoresistance (AMR) \cite{AMR_cam,AMR_kokado,kokado2013,AMR_Kokado_2015}.

AMR describes the change of the electrical resistance as a function of the angle between the electric current and the magnetization direction. Maximum resistance is observed when the direction of the current is aligned with the magnetization direction; minimum resistance is observed when the current and magnetization are perpendicular. The resistivity $\rho$
is given by \cite{De_Ranieri_2008}:
\begin{equation}
    \rho (\phi)=\rho _{0}+ \Delta \rho\rm{cos}^2 \phi
    \label{eqn:AMR}
\end{equation}
where $\phi$ is the in-plane angle between the electric current and magnetization, $\rho_0$ is the isotropic resistivity, and $\Delta \rho$ is the anisotropic resistivity change, which is defined as $\Delta \rho=\rho_{\parallel}-\rho_{\perp}$. Here, 
$\rho_{\perp}$, $\rho_{\parallel}$ represent the resistivity when $\phi$ is $90^{\circ}$ and $0^{\circ}$, respectively.

\begin{figure}
    \centering
    \includegraphics[width=\linewidth]{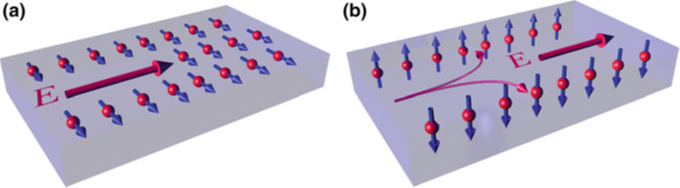}
    \caption{Illustration of (a) inverse spin galvanic effect and (b) spin Hall effect. The inverse spin galvanic effect creates a homogeneous spin density throughout the sample as shown in (a). In comparison, the spin Hall effect results in the accumulation of spins of opposite spin orientations on the lateral surfaces of the sample as shown in (b). {Reproduced with permission from M. B. Jungfleisch et al., \textit{Spin-Orbit Torques and Spin Dynamics in Spin Physics in Semiconductors} in Springer Series in Solid-State Sciences, Vol. 157 (Springer, Cham, 2017). Copyright 2017 Springer-Verlag Berlin Heidelberg.}}
    \label{fig:Spin_hall}
\end{figure}

\paragraph{Spin Hall effect}
The ordinary Hall effect is the observation of a voltage transverse to an electric current when a conductor is exposed to an external magnetic field that is perpendicular to the direction of the electric current \cite{Hall_effect}.
In the early 1970s, Dyakonov and Perel proposed that spins accumulate on the lateral surfaces of a conductor through which an electric current is passed, even in the absence of an external field; an effect that is now commonly referred to as the spin Hall effect \cite{Dyakonov1971}. Due to spin-orbit coupling in the conductor, electrons with opposite spin alignment will be scattered in opposite perpendicular directions relative to the flow of the electric current. The spin Hall effect is closely related to the anomalous Hall effect observed in magnetic materials; however, the spin Hall effect does not require magnetic ordering.
Phenomenologically, a description of creating transverse charge current from spin current is given by: 
\begin{equation}
    \Vec{j}_c \propto \theta_\mathrm{SHE} \Vec{j}_s \times \Vec{\sigma},
\end{equation}
where $\Vec{j}_c, \Vec{j}_s$ are the charge current density and spin current density, respectively, $\vec{\sigma}$ is the direction of the spin polarization, and $\theta_\mathrm{SHE}$ is the spin Hall angle of the material, which is a measure of the material-specific charge-to-spin conversion efficiency \cite{HoffmannIEEETM2013}{, see Fig.~}\ref{fig:Spin_hall}. The most commonly used spin-Hall materials are heavy metals with a strong spin-orbit interaction, such as Pt, Ta, and W. 
A typical spintronic THz emitter relies on the inverse of the spin-Hall effect (ISHE) \cite{HoffmannIEEETM2013}: here, the spin-Hall material converts an ultrafast spin current to a charge current transient that gives rise to THz radiation, e.g., Ref.~[\onlinecite{Kampfrath_Nat2013}], [\onlinecite{Walowski_JAP2016}].

\paragraph{Spin galvanic effect {/ inverse Rashba Edelstein effect}}

\begin{figure}
    \centering
    \includegraphics[width=0.7\linewidth]{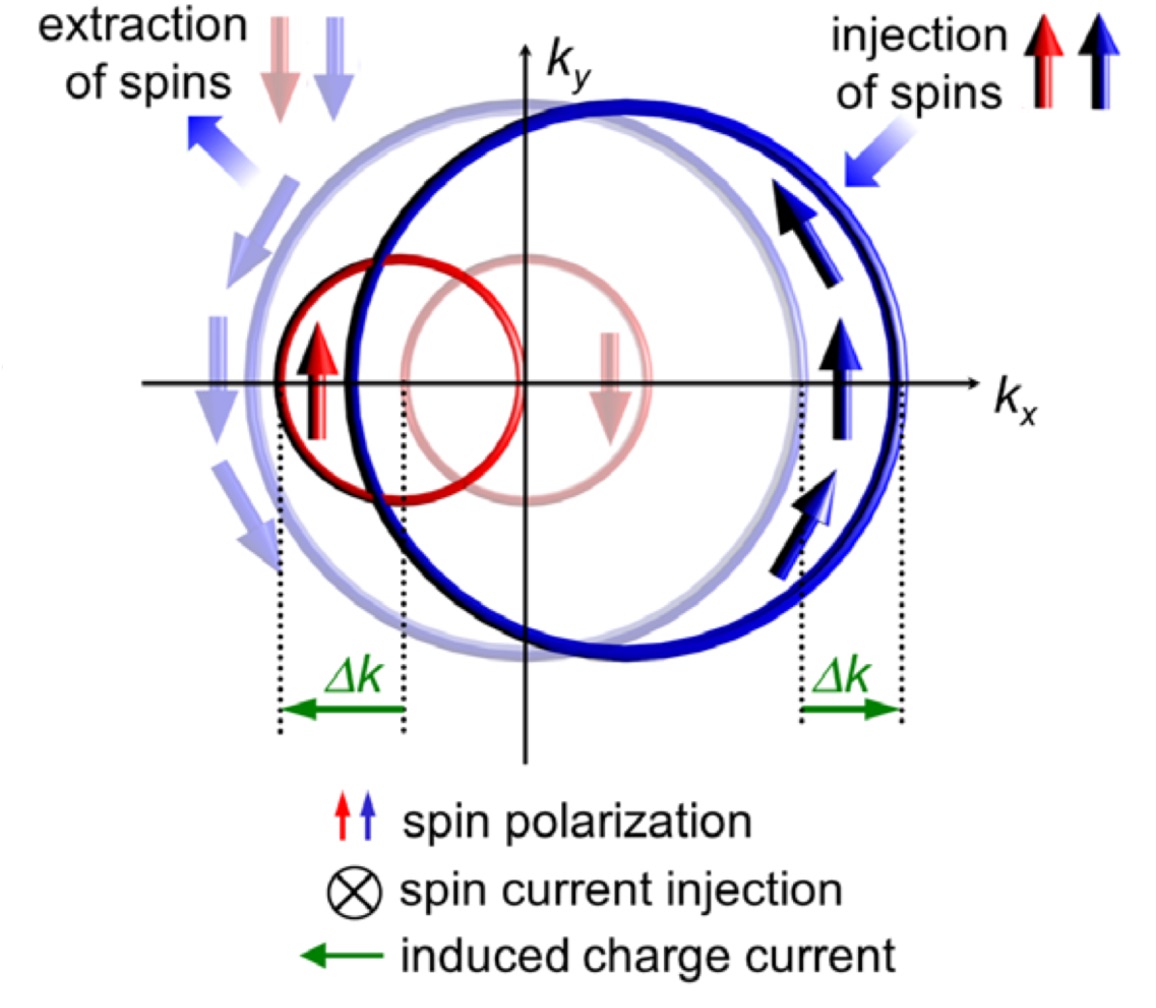}
    \caption{{Illustration of spin-to-charge current conversion by means of the inverse Rashba Edelstein effect: The generation of a charge current carried by the interfacial states is due to a nonzero spin density induced by spin-current injection. Reproduced with permission from Phys. Rev. Lett. \textbf{120}, 207207 (2018). Copyright 2018 American Physical Society.}}
    \label{fig:IREE_mechanism}
\end{figure}

The  spin galvanic effect {(sometimes also referred to as inverse Rashba Edelstein effect)} \cite{Ganichev_2002,Ganichev_2003,Ganichev_2004} is another mechanism for spin/charge interconversion \cite{Sanchez_Nat2013,ZhangJAP2015,Jungfleisch_PRB_Rashba_2016,Nakayama_PRL2016}. In contrast to the spin-Hall effect, the inverse spin galvanic effect (or Rashba Edelstein effect) describes the electric generation of a homogeneous spin density throughout the sample, rather than only on the conductors' surfaces. It relies on the asymmetric spin-flip scattering of electrons in systems where the dispersion of the two electron subbands is spin-split due to a broken inversion symmetry. We note that in general a broken inversion symmetry is not enough; the inverse spin galvanic effect can only be observed in gyrotropic materials \cite{Jungfleisch_Springer2017}. When an electric field is applied to the system, the Fermi contours shift, giving rise to a non-equilibrium steady-state spin polarization perpendicular
to the driving electric field. The inverse process is called the spin galvanic effect: here an electric charge current is created by spin-current injection. {This is schematically shown in} Fig.~\ref{fig:IREE_mechanism}. The aforementioned considerations also apply to interfaces, i.e., it is possible to artificially create layered structures of non-gyrotropic materials in the bulk which become gyrotropic when assembled as multilayers and thus the inverse spin galvanic effect can be observed in these systems \cite{Jungfleisch_Springer2017}. {Examples of multilayers exhibiting Rashba type interfaces that can be used for spin-to-charge conversion include Bi/Ag and Bi/Sb} \cite{Sanchez_Nat2013,ZhangJAP2015,Jungfleisch_PRB_Rashba_2016,Nakayama_PRL2016}.

{The spin galvanic effect (or inverse Rashba Edelstein effect) has been used to demonstrate effective spin-to-charge conversion by steady-state spin pumping \cite{Sanchez_Nat2013,ZhangJAP2015} and spin-torque ferromagnetic resonance \cite{Jungfleisch_PRB_Rashba_2016}. It has also been demonstrated in magnetoresistance measurements \cite{Nakayama_PRL2016,Jungfleisch_PRB_Rashba_2016}. Recently, the spin galvanic effect was used in the ultrafast regime showing that it can even be utilized in spintronic THz emitters \cite{Jungfleisch_PRL2018,Zhou_PRL2018,Li_PRMat2019}.}

\begin{figure*}[ht]
    \centering
    \includegraphics[width=0.99\linewidth]{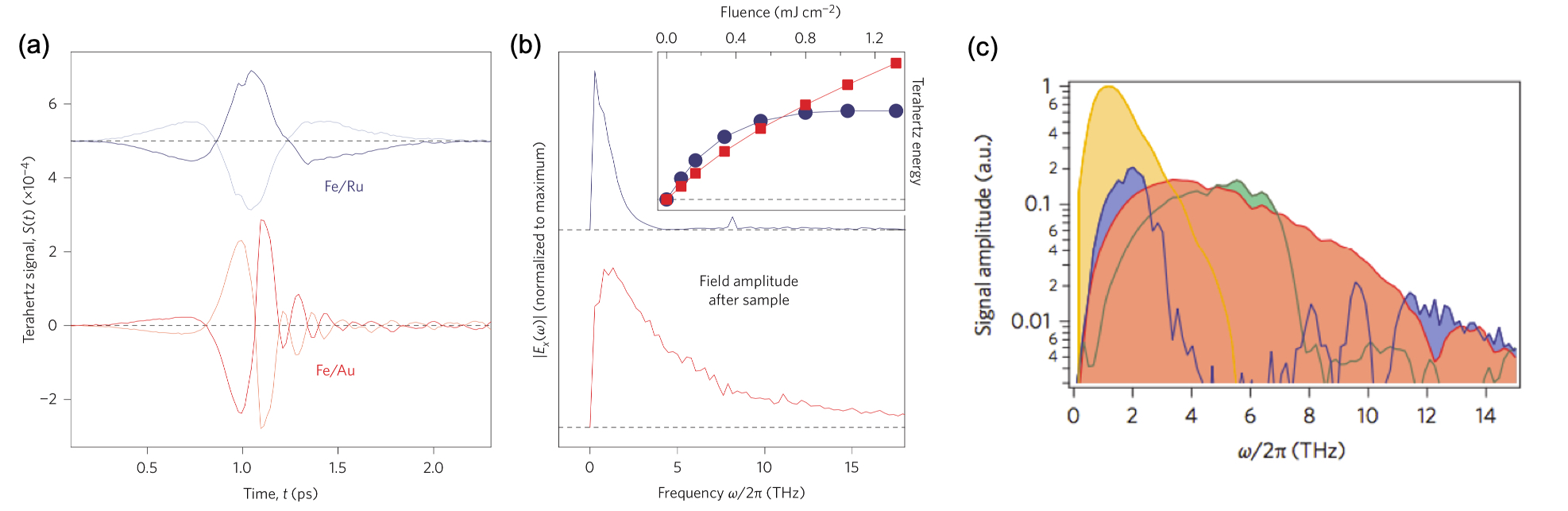}
    \caption{Terahertz emission from spintronic heterostructures. (a) THz signal signal trace obtained from photoexcited Ru- and Au-capped Fe thin films. The signal inverts when the sample magnetization is reversed (dark to light curves). (b) Corresponding Fourier spectra. Inset: emitted THz pulse energy vs absorbed pump-pulse energy per area. {Reproduced with permission from Nature Nanotechnology \textbf{8}, 256-260 (2013). Copyright 2013 Springer Nature.} (c) Spectrum of the spintronic W/Co$_{40}$Fe$_{40}$B$_{20}$/Pt trilayer emitter in comparison to standard terahertz emitters measured with a 70-$\mu$m-thick Lemke/amorphous polycarbonate electrooptic sensor (measurements performed under identical conditions, thicknesses given in parentheses). Orange: photoconductive switch, blue: ZnTe(110) (1 mm), green: GaP(110) (0.25 mm), red: spintronic emitter (5.8 nm). {Reproduced with permission from Nature Photonics \textbf{10}, 483-488 (2016). Copyright 2016 Springer Nature.}}
    \label{fig:Kampfrath}
\end{figure*}

\section{Design, synthesis and fabrication of spintronic terahertz emitters}\label{SpintronicTHzDevices}

A key advantage of spintronic THz emitters is their relatively easy synthesis and fabrication. Spintronic devices are usually fabricated by thin-film deposition techniques such as sputter deposition, e-beam evaporation, or molecular beam epitaxy (MBE). The same standard techniques can be employed for the synthesis of spintronic THz emitters. While most work to date relies on the fast and inexpensive sputtering technique, recent efforts have used molecular beam epitaxy to understand the influence of electron-defect scattering lifetime, structural defects, and the substrate material on the THz properties \cite{Torosyan_2018,Nenno_2019}. 

In the following, we review what material properties limit the performance of spintronic THz emitters and how various synthesis methods and strategies can be used to address these shortcomings. In general, the pulse length, and thus the THz bandwidth, are determined by the temporal characteristics of the charge current transient induced by the inverse spin-Hall effect. The rising edge of the pulse is dictated by the optical pump pulse duration and the spin diffusion properties. The spin diffusion properties depend on the density of states, band velocity, and scattering rate of electrons when they are promoted above the Fermi level in the FM \cite{Seifert_Nat2016} and on the spin transmission across the FM/NM interface. The falling edge of the pulse is determined by the carrier relaxation times in the NM layer \cite{Torosyan_2018}. The magnitude of the ultrafast spin-to-charge current conversion, which largely determines the THz pulse amplitude, are determined by the spin-Hall angle, spin diffusion length, resistivity, and interfacial spin transport properties. 

Various approaches and experimental strategies have been explored to modify THz emission characteristics. Initial work mainly focused on the exact material composition and thickness of the FM/NM sample stack, including NMs with different magnitude and sign of the spin-Hall angle and thickness-dependent Fabry-P\'erot-type resonances \cite{Kampfrath_Nat2013,Seifert_Nat2016}. More recent efforts have focused on new material components such as ferrimagnets \cite{Schneider2018,Chen_2018}, antiferromagnets \cite{Chen_2018}, and synthetic antiferromagnets \cite{Qi_PRApplied2020}. Continued study of the typical FMs \cite{Chen_2018,Seifert_SPIN2017,Schneider2018,Huisman_2017,Qi_PRApplied2019} has focused on the influence of the interface quality on the THz emission \cite{Li_PRMat2019,Li_2018,Seifert_2018JPDAP} and the role of the crystal growth and crystallinity of the materials \cite{Seifert_Nat2016,Sasaki_APL2017, Torosyan_2018, Nenno_2019}. In Sect.~\ref{sec:SpintronicCrystallinity} we focus on how interface and crystal quality affect THz emission properties. In Sect.~\ref{sec:sources}, we review the different ``building blocks'' (materials, thickness, magnetic ordering) of spintronic THz emitters. We stress that we are just at the beginning of this field and there is tremendous opportunity to exploit ``conventional spintronics" knowledge of transport, interfacial effects, etc.~at both DC and GHz frequencies to engineer much more sophisticated heterostructures and interfaces to optimize performance at THz frequencies.

\subsection{Impact of interface and crystal quality on THz emission}\label{sec:SpintronicCrystallinity}
Torosyan et al.~presented systematic studies of how THz emission from MBE-grown Fe/Pt bilayers on MgO and sapphire substrates depends on layer thickness, growth parameters, substrates, and geometrical arrangement \cite{Torosyan_2018}. They found that Fe/Pt bilayers on MgO substrates have a higher dynamic range below 3 THz when compared to bilayers on Al$_2$O$_3$ substrates. This difference was attributed to the almost-epitaxtial growth of Fe on the MgO substrates. By varying Fe and Pt layer thicknesses, they found that samples with 2 nm Fe and 3 nm Pt resulted in the maximum THz emission amplitude. The experimental results were supported by a model that takes into account the generation and diffusion of hot electrons in the Fe layer, the shunting effect, spin accumulation in the Pt layer, and optical properties of the bilayers\cite{Torosyan_2018}. Torosyan et al.~also employed a Si lens to collimate the THz beam and showed that the best result can be obtained when the lens faces the substrate side and not the metal layer. This can be understood by a suppression of the reflections at the MgO/air interface and has the additional advantage that the delicate Pt surface is not damaged when the lens is mounted. A follow-up work by the same group investigated the influence of the electron-defect scattering lifetime on the spectral shape and how structural defects affect the interface transmission and thus the THz amplitude \cite{Nenno_2019}.

As noted above, not only the spin-diffusion properties and the spin-Hall angle, but also the interface transmission properties are critically important. Li et al.~presented a detailed investigation of the effects  on the THz emission of roughness, interface intermixing, and crystal structure \cite{Li_PRMat2019} in Co/Pt heterostructures. For this purpose, polycrystalline samples with different roughness were grown by magnetron sputtering, using control over the deposition chamber pressure to alter the microstructure at the interface. Li et al.~also fabricated Co/Co$_\mathrm{x}$Pt$_\mathrm{1-x}$/Pt samples to study the influence of the intermixing of Pt and Co atoms at the interface and compared the results obtained from polycrystalline samples to those obtained from eptiaxtial samples. In essence, they found that the photocurrents created by the helicity-independent THz emission due to the inverse spin-Hall effect and the helicity-dependent THz emission due to the spin-dependent photogalvanic effect (or spin-galvanic effect) \cite{Ganichev_2002} show opposite trends: the helicity-independent ISHE contribution decreases as the roughness is increased, while the helicity-dependent contribution increases from near zero as the roughness increases. The former effect is explained by a suppression of the spin-current transparency across the interface due to roughness-induced enhanced spin-flip probability. The latter observation is explained by a geometrical increase of the bulk volume that ``feels'' the interface properties responsible for the spin-dependent photogalvanic effect. These measurements also revealed that intermixing enhances the helicity-independent contribution of the THz emission, but suppresses the helicity-dependent contribution. Li et al.~propose that the effects of intermixing are due to an enhanced spin-current transmission across the interface and/or a higher spin-orbit coupling in the Co$_\mathrm{x}$Pt$_\mathrm{1-x}$ layer. Finally, Li et al.~found that epitaxially-grown samples do not show any helicity-dependent emission.

We note that Jaffr\'es et al.~correlated the ferromagnetic-resonance driven spin-pumping results with THz emission from optimized 3D/5D heavy metal interfaces. They found a strict correlation between THz signals and the spin-mixing conductance and ISHE signals \cite{Jaffres_2019,Dang_2020}. {Furthermore, Gueckstock et al. systematically studied the effect of the ferromagnet - normal metal interface and found dramatic changes in the amplitude and even an inversion of the polarity of
the THz emission. These results suggest that the interfacial spin-to-charge current conversion arises from skew scattering of spin-polarized electrons at interface imperfections} \cite{Gueckstock_2021}.



\subsection{Terahertz generation based on the inverse spin Hall effect and spin Seebeck effect}
\label{sec:sources}
Kampfrath et al.~were the first to report an ultrafast, contactless amperemeter based on a spintronic THz emitter that utilized the inverse spin Hall effect to convert an ultrafast spin flow into a terahertz electromagnetic pulse \cite{Kampfrath_Nat2013}. Since that discovery, a variety of different materials have been explored as potential spin-based THz sources. Those studies not only helped to optimize and engineer the THz signal, but also contributed to our current understanding of ultrafast magnetic phenomena in magnetic heterostructures and multilayers. In this section, we present a summary of some pioneering work that relied on spin-to-charge conversion by means of the inverse spin Hall effect in both metallic and insulating magnets.
Terahertz sources based on metallic heterostructures have key advantages. Their fabrication is well established, inexpensive, and can easily be scaled up using large-scale physical vapor deposition techniques such as sputtering deposition. This manufacturing advantage makes metallic sources attractive inexpensive alternatives to the commonly used THz sources based on PCAs and nonlinear crystals, which were described in Sect.~\ref{sec:traditional_sources}. The absorption in metallic sources is to a large degree independent of the pump wavelength and they feature a very short electron lifetime (down to 10 fs) enabling a broadband emission covering a frequency range between 1 and 30 THz \cite{Seifert_Nat2016}. In the following, we discuss the state of the art beginning with the pioneering work that combined concepts of spintronics with ultrafast magnetism.

\textbf{First observation.} In 2013, Kampfrath et al. discovered the possibility to control ultrafast spin current pulses by magnetic heterostructures made of ferromagnetic iron thin films and nonmagnetic capping layers (Ru and Au)\cite{Kampfrath_Nat2013}. The observed THz emission from the magnetic heterostructures was interpreted to arise from a photo-excited spin-polarized electron current created in the iron layer and the subsequent conversion into charge current transients in the the nonmagnetic layer due to the inverse spin Hall effect. By absorbing a femtosecond laser pulse, electrons from below the Fermi level are promoted to bands above the Fermi energy creating ``hot'' electrons. The non-equilibrium hot majority electrons in iron are sp-type electrons and have a higher velocity than the d-type minority electrons, resulting in an ultrafast spin-polarized electron current that is created in iron and injected into the capping layer material. Ru and Au are chosen for the capping layer due to their very different electron mobility leading to different transport dynamics. In Au, the hot  electrons occupy sp bands with a high velocity and long lifetime, while in Ru they occupy d bands with a lower velocity and have a shorter lifetime. As a result, Kampfrath et al. found different THz waveforms from the photoexcited Ru- and Au-capped iron films, Fig.~\ref{fig:Kampfrath}(a,b). This pioneering work paved the route towards THz spintronic devices and applications. It also provided new insights into the underlying mechanisms of ultrafast spin physics and introduced a new method to contactlessly detect spin current in the THz frequency regime.

\begin{figure}[t]
    \centering
    \includegraphics[width=\linewidth]{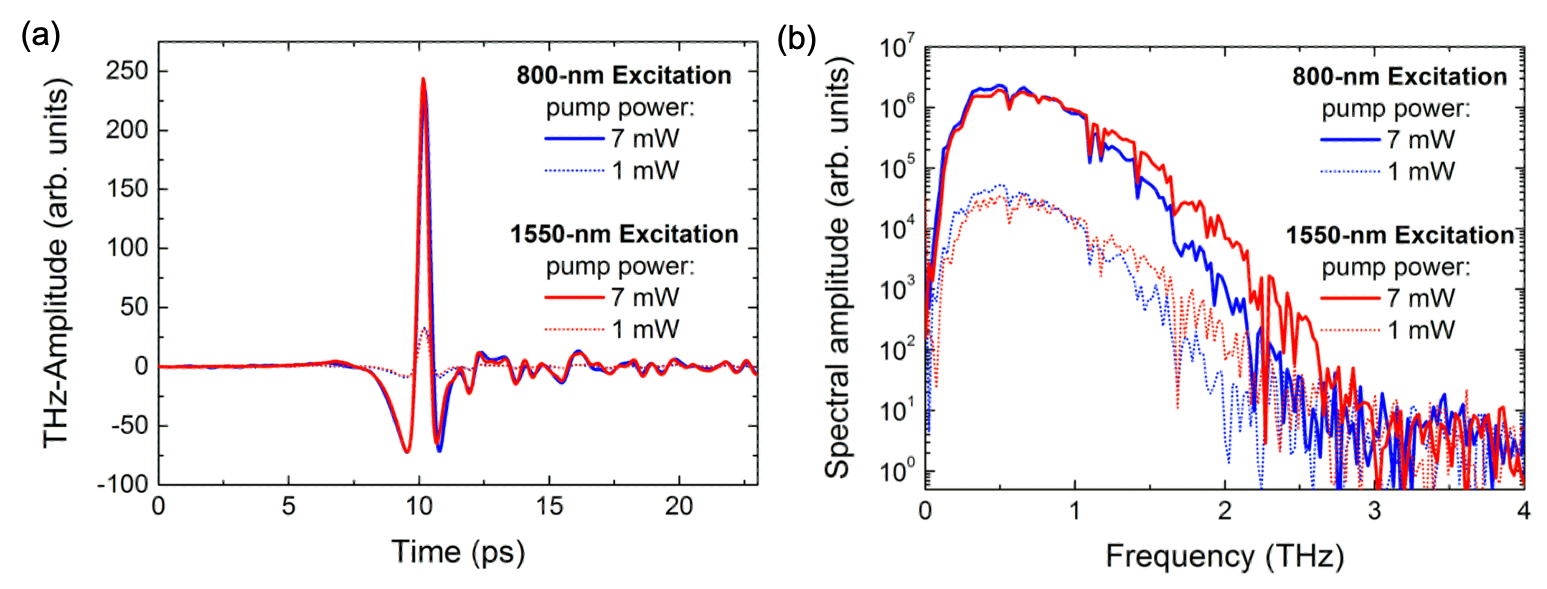}
    \caption{(a) THz waveforms and (b) corresponding spectra obtained using two laser excitation wavelengths (800 and 1550 nm) for a Fe/Pt bilayer structure. {Reproduced with permission from IEEE Transactions on Magnetics \textbf{54}, 9100205 (2018). Copyright 2018 IEEE.}}
    \label{fig:wavelength}
\end{figure}

\textbf{Optimized efficiency of metallic spintronic THz emitters.}
Seifert et al.~presented an approach to optimize the bandwidth, amplitude, and scalability of spintronic THz emitters\cite{Seifert_Nat2016}. One fundamental advantage of spintronic emitters relative to III-V emitters is the absence of gaps in the emission spectra arising from interaction with phonons. This phonon interaction is forbidden in the generation of ultrafast photoexcited spin currents or in the inverse spin-Hall effect. Seifert et al.~employed two heavy metals with opposite spin-Hall angle and sandwiched the ferromagnetic layer between the heavy metal layers to enhance the spin-to-charge current conversion\cite{Seifert_Nat2016}. They also utilized a broadband Fabry–P\'erot resonance to further increase THz signal output\cite{Seifert_Nat2016}. Figure~\ref{fig:Kampfrath}(c) shows a comparison of the THz amplitude from traditional THz sources and the conceptually-different spintronic emitters. The spectrum obtained from the optimized sample stacks impressively shows the advantages of the spintronic emitters: the spintronic emitter made of W/CoFeB/Pt exceeds by far the bandwidth of the traditional emitters. Moreover, an intense broadband THz emission from metallic spintronic thin-film stacks composed of a W/CoFeB/Pt trilayer was reported in 2017\cite{Seifert2017}, with the peak field reaching 300~kV/cm, and a pulse energy of 5 nJ. As we will discuss in more detail in Sec.~\ref{sec:opportunities}, spintronic emitters are largely based on thin film deposition techniques, which means their emission can be further optimized using standard microfabrication techniques such as photo- and electron-beam lithography\cite{Yang_AOM2016,Lendinez_SPIE_2019,Weipeng_Wu_JAP2020}.

\textbf{Demonstration of insensitivity to laser pump wavelength.}
As discussed in Sec.~\ref{Sec:PCA}, semiconductor-based THz emitters (PCAs) require a certain laser pump wavelength. In contrast, it was recently demonstrated that metallic spintronic THz emitters are insensitive to the excitation laser wavelength \cite{Papaioannou_TMAG2018,Herapath_2019}. Herapath et al. reported that the efficiency of THz generation is independent of the pump-photon energy within central wavelengths ranging from 900 to 1500 nm \cite{Herapath_2019}. Papaioannou et al.~used two different laser wavelengths and compared the resulting THz emission from Fe/Pt bilayers \cite{Papaioannou_TMAG2018}. Figure~\ref{fig:wavelength} shows that the excitation of non-equilibrium spin-polarized electron currents with less photon energy ($\lambda= 1550$~nm) is essentially as effective as the use of higher photon excitation energy ($\lambda= 800$~nm) \cite{Papaioannou_2018}. However, at first glance one would expect the opposite observation: higher photon energies would lead to a larger asymmetry of the two spin species because they are residing in different bands with different band velocities. This perceived discrepancy can be understood by taking into account not only the directly excited high-energy electrons, but also secondary electrons at intermediate energies created due to electron–electron scattering events after the initial excitation of the  electrons. The lifetime decreases with the energy of the excited electron, and therefore the most energetic electrons do not significantly contribute to the process. The most significant contribution comes from intermediate energy electrons resulting from scattering, whose contribution is similar to that of electrons with longer lifetimes that are directly excited by lower-energy photons \cite{Papaioannou_TMAG2018}. Note that at very low photon energies we also have to consider the contribution of holes to the spin-diffusion process, which results in a zero net transport of spin \cite{Papaioannou_TMAG2018}. Figure~\ref{fig:wavelength} shows that the total energy that it is deposited into the system is the important quantity here. 

While the excitation laser wavelength is not directly critical to the THz generation, Herapath et al.~demonstrated that including TiO$_2$ and SiO$_2$ dielectric overlayers after the optimization for a particular excitation wavelength can further enhance the terahertz emission \cite{Herapath_2019}.


\textbf{Insulator-based spintronic THz emitters.}

Very recently, it was shown that THz emission can also be observed in magnetic insulator-based heterostructures upon photo-excitation \cite{Seifert2018, Cramer2018}. In these studies the magnetic insulator yttrium iron garnet was heated by an adjacent metallic layer resulting in a spin-Seebeck effect-driven spin current that arises on a time scale of about 100 fs \cite{Seifert2018}. This time scale is comparable to the time scale of the thermalization process of the metal electrons.
These studies provide important insights into the spin transfer and the fundamental time limitations for angular-momentum transfer across metal-insulator interfaces. We note, however, that the signal strength is considerably smaller than for metallic spintronic sources. Consequently, we do not further discuss insulator-based spintronic THz emitters and refer the reader to the literature \cite{Seifert2018,Cramer2018}.




\subsection{Terahertz generation based on spin galvanic effect}

While most work relies on the inverse spin Hall effect in the bulk for converting ultrafast photo-excited spin currents into charge current transients, there has been some effort in exploring interfaces. Employing interfaces offers important advantages including effects such as interface-induced magnetism and non-trivial spin textures, spin-transfer torque effects, and magnetization reversal induced by interfaces \cite{Hellman_RevMod_2017}. Furthermore, potential devices could be made thinner and thus require less material and a smaller volume, interfaces are easier to manipulate and even gateable \cite{Lesne_NatMat_2016}, and layered structures with multiple interfaces could potentially be used to increase the overall effect. Efficient spin-to-charge conversion at interfaces between two dissimilar materials was first demonstrated in steady-state spin pumping experiments at GHz frequencies \cite{Sanchez_Nat2013,ZhangJAP2015}, in spin-torque ferromagnetic resonance \cite{Jungfleisch_PRB_Rashba_2016}, and in magnetoresistance measurements \cite{Nakayama_PRL2016}. 
Recently, it was shown that spin-to-charge conversion at Rashba interfaces can also be used on ultrafast time scales \cite{Huisman_Nat2016,Jungfleisch_PRL2018,Zhou_PRL2018,Li_PRMat2019}. 

\begin{figure}[t]
    \centering
    \includegraphics[width=\linewidth]{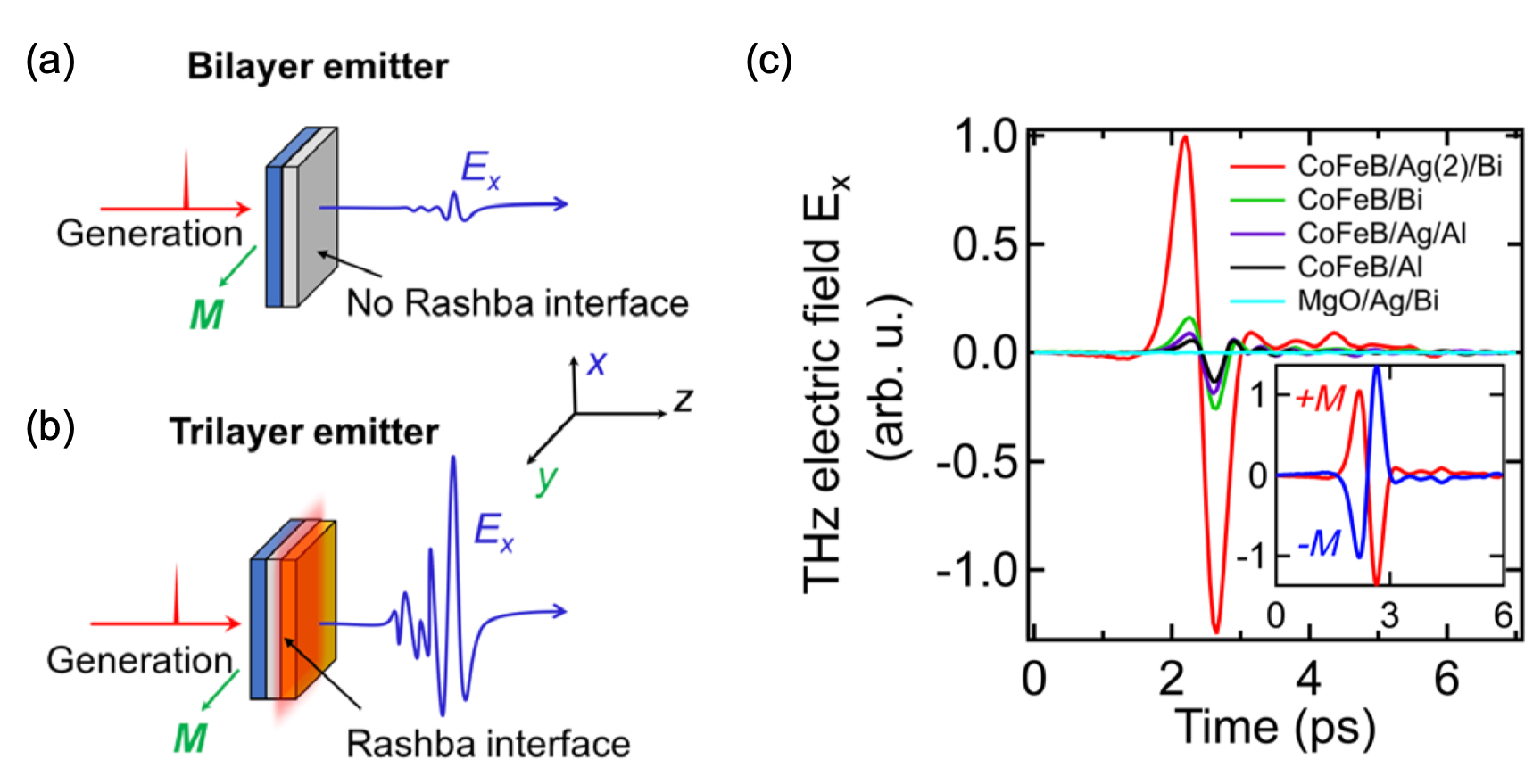}
    \caption{(a) Experimental configuration and sketch of THz emission characteristics of (a) control samples without Rashba interface (CoFeB/Bi, CoFeB/Ag/Al, CoFeB/Al, MgO/Ag/Bi) and (b) trilayer sample (CoFeB/Ag/Bi), where the spin-to-charge conversion occurs at the Rashba interface between Ag and Bi. (c) Comparison of the CoFeB/Ag/Bi trilayer with bilayer control samples. An enhancement of the signal strength is observed when Bi is deposited onto Ag. The inset shows that the signal is inverted when the magnetization $M$ is flipped. Reproduced with permission from {Phys. Rev. Lett. \textbf{120}, 207207 (2018). Copyright 2018 American Physical Society.}}
    \label{fig:IREE}
\end{figure}

There are a number of reports focused on the Rashba interface between Ag and Bi for spin-to-charge conversion. In particular, it was shown that the THz emission can be enhanced by a factor of 6 when a bilayer of Ag/Bi is deposited onto the metallic ferromagnet CoFeB \cite{Jungfleisch_PRL2018}. Furthermore, a helicity dependence of the THz electric field that is polarized parallel to the CoFeB magnetization direction was observed. This effect was absent in any of the control measurements on samples that do not exhibit the Ag/Bi interface and is most pronounced for thin Ag layers below 10 nm \cite{Jungfleisch_PRL2018}. Similarly, it was shown that Fe can serve as a spin current source material adjacent to a Ag/Bi bilayer\cite{Zhou_PRL2018}. 



\section{Outlook and perspective}\label{sec:outlook}
\subsection{Rise of antiferromagnetic spintronics}



{Recent work on ultrafast spintronic effects in antiferromagnetic and ferrimagnetic materials revealed interesting properties and ultrafast phenomena distinctly different from their ferromagnetic counterparts}\cite{Finley_APL2020,Gomonay_2017,Jungfleisch_PLA2018}. {In antiferromagnets the sublattice magnetizations may oppose each other, while in compensated ferrimagnets the net magnetization can be tuned by composition, temperature, or strain} \cite{Finley_APL2020}. 
{Antiferromagnets exhibit strong exchange coupling between neighboring spins; much stronger than in ferromagnets} \cite{Qiu2020}{. Therefore, antiferromagnets show intrinsic ultrahigh resonance frequencies in the THz range, whereas ferromagnets have resonances in the GHz range} \cite{Baltz_AFM, Jungfleisch_PLA2018}.
{Utilization of materials with magnetic order but vanishing net magnetization offers numerous advantages over their ferromagnetic counterparts. These advantages include a higher stability due to the absence of stray fields, higher operational speed, resonance frequencies in the THz range, and the existence of both counter-clockwise and clockwise spin-wave modes} \cite{Jungfleisch_PLA2018,Gomonay_2017}. {The detailed investigation of these systems in terms of their ultrafast spintronic properties started very recently and there are many open questions that demand further study. In the following we focus our discussion on ferrimagnets and non-collinear antiferromagnets. In particular, we will review the THz emission characteristics of heterostructures consisting of ferrimagnets and non-collinear antiferromagnets and describe their prospects as future spintronic THz devices.}

{Most recent studies} on the spin dynamics of antiferromagnets have focused on collinear antiferromagnetic spin alignments. A typical antiferromagnetic insulator is NiO, which has resonances in the low THz range. In particular, recent THz spectroscopy studies of NiO revealed the damping parameter \cite{moriyama_enhanced_2020,moriyama_intrinsic_2019} and the magnetic field and temperature dependence of the antiferromagnetic resonance \cite{wang_magnetic_2018}. {Wang} et al. reported magnon-torque–induced magnetization switching and THz emission in Bi$_2$Se$_3$/ NiO/NiFe devices at room temperature \cite{Magnetization-switching-Wang1125}{; this effect was later modeled by Suresh et al. using quantum-classical simulations based on a time-dependent nonequilibrium Green functions combined with the Landau-Lifshitz-Gilbert equation framework, which confirmed that the magnon torque plays the dominant role in the system} \cite{Suresh}. {Qiu et al. observed THz emission from NiO/Pt and NiO/W heterostructures at zero external magnetic field and at room temperature}\cite{Qiu2020}. {In particular, they observed a connection between the angular-dependent THz waveforms and the three-fold rotational symmetry of NiO, which suggest the possibility to 1) infer symmetry properties of the system from its THz emission characteristics and 2) control THz emission through structural symmetry or propagation direction.}  


Chen et al. \cite{chen_terahertz_2018} reported that the THz emission from transition metal-rare earth element ferrimagnet/heavy metal CoGd/Pt bilayer structures depends strongly on both temperature and chemical composition. Interestingly, strong terahertz emission from this structure with nearly zero net magnetization is observed. The authors explain the results by the localized and delocalized nature of the electrons carrying magnetic moments in Gd and Co, respectively. For generating the superdiffusive spin current in the ferrimagnetic layer, 3d-electrons from the Co sublattice can be excited more easily than the 4f-electrons from the Gd sublattice, which are further below the Fermi level. In the same work, the THz emission measurements of antiferromagnetic IrMn-based heterostructures show that no measurable THz signal is observed in the IrMn/Pt bilayer. However, when the antiferromagnet is used as a spin detector in a IrMn/Co bilayer -- where Co is a metallic ferromagnet -- a sizable THz signal is detected, indicating that antiferromagnets can be used a spin-to-charge converter even at ultrafast timescales. Indeed, it was previously reported in steady-state spin pumping measurements at GHz frequencies that Mn-based antiferromagnets show a sizeable spin-Hall angle and hence can be used as spin current detectors \cite{zhang_spin_2014,ZhangPRB2015,MendesPRB2014}. 

Meanwhile, Schneider et al. \cite{schneider_magnetic-field-dependent_2018} studied another transition metal-rare earth element ferrimagnet, FeTb. Schneider et al. found that  the THz emission amplitude from a TbFe/Pt layer closely follows the in-plane TbFe magnetization. 
This is in contrast to the previously discussed results obtained for CoGd, where THz emission was observed even for a zero net magnetization and presumed to be due to the difference in the transition metal-rare earth element bandstructure:  The electrons of the Gd sublattice that can be excited by the ultrafast laser pulses are localized 8 eV below the Fermi level in the 4f-band. Therefore, the ferromagentic Co sublattice is the dominant contribution to the spin-polarized current generated upon excitation with the femtosecond laser pulse. In contrast, in the Tb sublattice, the excited electrons arise from the more-than-half-filled 4f shell, which is about 2.23 eV below the Fermi level. Furthermore, in TbFe the hybridization of 3d, 4f, and 5d electrons leads to a situation in which it is easier to excite electrons from the Tb sublattice than from the Gd sublattice (while still being more difficult than exciting the Fe sublattice). Therefore, the THz emission from FeTb closely follows the in-plane magnetization and a complete cancellation of THz emission is observed.



\begin{figure*}[t]
    \centering
    \includegraphics[width=0.7\linewidth]{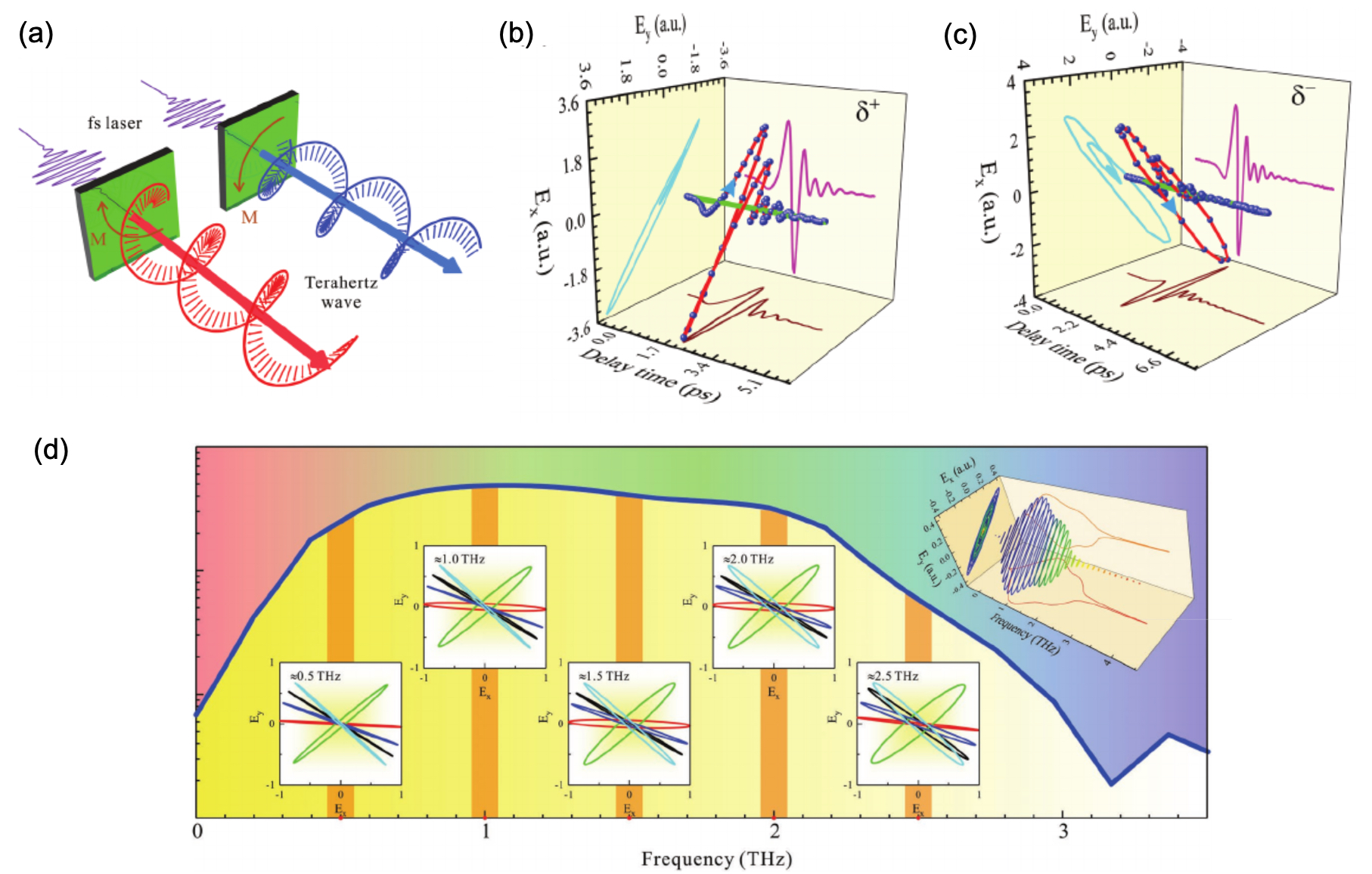}
    \caption{(a) Conceptual illustration of the terahertz chirality control using a magnetic field profile. (b) Experimentally observed parameteric plots of the THz waveforms with left-handed and right-handed elliptical polarizations. (d) Broadband elliptical THz emission. The colors represent different polarization states. Black: linear polarized ($\delta^\circ$); red, green: right-handed ($\delta^-$); blue, light blue: left-handed ($\delta^+$) polarization. Inset: 3D broadband polarization spectrum of a right-handed elliptically polarized THz wave. 
    {Reproduced with permission from Adv. Optical Mater. \textbf{7}, 1900487 (2019). Copyright 2019 John Wiley and Sons.}}
    \label{fig:inhomogeneous}
\end{figure*}

More recently, non-collinear antiferromagnets have attracted increased attention due to their interesting spin configurations. In particular, it was shown that non-collinear antiferromagnets can be employed for generating and controlling the THz emission in antiferromagnetic spintronic THz emitters. 
It was revealed that heterostructures based on the non-collinear antiferromagnet Mn$_3$Sn \cite{zhou2019orientation} and excited by a femtosecond laser pulse can either be used as a spin current source (Mn$_3$Sn/Pt) or serve as a spin-to-charge converter when combined with a ferromagnetic metal (Co/Mn$_3$Sn). In the latter case, the anisotropic inverse spin Hall effect is responsible for the conversion. 
Finally, Zhang et al. reported THz radiation from an exchange-coupled synthetic antiferromagnet where two antiparallelly aligned ferromagnetic layers are separated by a spacing layer \cite{Synthetic_AFM}.






\subsection{New opportunities enabled by patterned terahertz sources and nonuniform magnetization textures}
\label{sec:opportunities}
Easy manipulation of the THz wave polarization and spectral bandwidth is important for next generation functional on-chip THz emitters. Spintronic THz sources are particularly interesting in this regard. Micro- and nanofabrication of magnetic heterostructures is readily available for patterning arrays of spintronic THz sources. Moreover, the magnetization state can be easily controlled by the externally applied magnetic field, offering conceptually new mechanisms for next generation THz applications and devices.

Yang et al.~presented detailed studies of the THz emission from Fe/Pt heterostructures \cite{Yang_AOM2016}. They not only investigated the thickness dependence of the bilayer and compared the THz waveforms to photoconductive switches and nonlinear crystals, but also demonstrated the tunability of the THz spectrum of patterned magnetic heterostructures via an external magnetic field. Lendinez et al. studied the effect of the stacking order on the THz waveform from CoFeB/Pt stripes \cite{Lendinez_SPIE_2019}. They found that the THz electric field amplitude is proportional to the coverage of the CoFeB/Pt heterostructure on top of the MgO substrate. Furthermore, by comparing the experimental results to micromagnetic simulations, they revealed that a small portion of the moments at the edges of the micron-sized stripes lie in the easy axis independent of the applied magnetic field, reducing the overall spin current that contributes to the THz signal. 
Wu et al.~showed a systematic dependence of THz emission characteristics on the size of Fe/Pt
bilayer stripes \cite{Weipeng_Wu_JAP2020}. These experimentally-observed spectra were interpreted in terms of a simplified multi-slit interference model that was capable of capturing the main experimental features.

In addition to using patterned magnetic heterostructures for manipulating the magnetization state and thus the THz emission properties, there has been an effort to control the local magnetization texture using inhomogeneous external magnetic fields \cite{Kong2019,Hibberd_2019}. Such magnetoelectric control of the magnetization may be another avenue for THz control in spintronic sources. Kong et al.~demonstrated the magnetic-field-controlled switching of the THz polarizations between linear and elliptical states in W/CoFeB/Pt trilayers using non-homogeneous field-induced magnetization states, as shown in Fig.~\ref{fig:inhomogeneous}. {The direction of the THz transient is controlled by the direction of the spin polarization vector as stated by the inverse spin Hall effect. By achieving a clockwise or counterclockwise spin polarization vector over the sample area using inhomogeneous magnetic field, they successfully tuned the direction of the converted charge current in Pt and W layer by the inverse spin Hall effect, resulting in elliptically polarized THz pulses} \cite{Kong2019}. Furthermore, the chirality, azimuthal angle, and ellipticity of the generated elliptical THz signal were tuned by applying an  external magnetic field. Similarly, Hibberd et al.~presented a proof-of-principle concept that a specific magnetic field distribution can be applied such that the transverse polarization of the resulting THz wave can be controlled \cite{Hibberd_2019}. By placing the spintronic emitter between two magnets of opposing polarity, a quadrupole-like polarization profile of the THz wave was observed. Furthermore, an enhanced longitudinal electric field component was observed in this configuration \cite{Hibberd_2019}.

\subsection{Integration of semiconductors}

\begin{figure}[t]
    \centering
    \includegraphics[width=\linewidth]{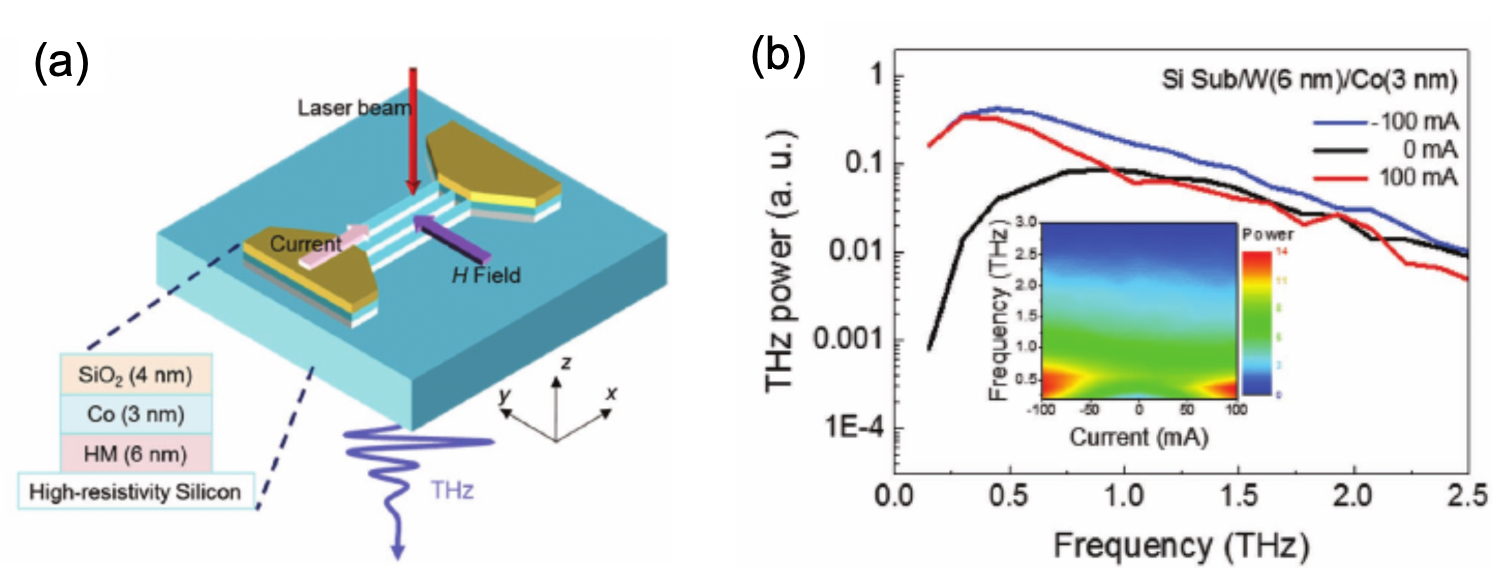}
    \caption{(a) A schematic depiction of a hybrid semiconductor / spintronic THz emitter. (b) The measured spectrum obtained with such a hybrid detector. 
     {Reproduced with permission from Adv. Optical Mater. \textbf{7}, 1801608 (2019). Copyright 2019 John Wiley and Sons.}}
      \label{fig:Si-W-Co}
\end{figure}

For spectroscopy applications it is desirable to have a THz source that emits an intense signal over the maximum possible bandwidth. The range from 0.1 to 10 THz is of particular interest for applications\cite{chen2019current}. As we have discussed in the preceeding sections, most spintronic emitters have relatively poor performance at frequencies below 1 THz and relatively strong performance at higher frequencies. Photoconductive antennas, on the other hand, tend to perform better at frequencies below 4 THz \cite{DreyhauptAPL2005, KlattOpticsExpress2009, GlobischJAP2017}. To generate relatively intense emission over a wider frequency range, Chen et al.~fabricated a hybrid emitter that combined a semiconductor PCA with a magnetic heterostructure\cite{chen2019current}. A schematic of this hybrid emitter is shown in Fig.~\ref{fig:Si-W-Co}(a). The PCA was made of HR-SI and the spintronic emitter was made of a either a Co/Pt or a Co/W bilayer. The spintronic bilayer had a total thickness less than 10 nm, far below the wavelength of THz radiation, and thus the emission of the PCA and spintronic components were effectively simultaneous. The data demonstrating the performance of the hybrid emitter with a Co/W bilayer is shown in Fig.~\ref{fig:Si-W-Co}(b). The black curve shows the spectrum obtained when no bias is applied to the PCA [Fig.~\ref{fig:Si-W-Co}(b)], effectively turning off the emission from the semiconductor component. In this case, only the spintronic component of the hybrid emitter is ``active''. As expected when only the spintronic emitter is contributing, the emission intensity at frequencies below about 1 THz is relatively poor. When either a positive or negative bias current (100~mA or $-100$~mA) is applied (red and blue curves, respectively), the semiconductor PCA  contributes to the emission and the THz signal at frequencies below 1 THz is significantly stronger. The difference in the intensity of the observed THz emission between the positively- and negatively-biased conditions originates in whether the PCA and spintronic THz emission interfere constructively or destructively, an observation that was confirmed by repeating the experiments with a Co/Pt bilayer. Because Pt has a spin-Hall angle with opposite polarity to that of W, the bias that resulted in constructive interference and enhanced THz emission for Co/Pt was opposite to that of Co/W.

The work of Chen, et al.~is an important first step toward the fabrication of hybrid THz emitters combining semiconductor and spintronic sources. The results also illustrate the range of opportunities that exist when creating hybrid materials. For example, one might choose to use two different wavelengths of light to excite the semicondutor and spintronic emitters at slightly different times. Another option may be to change the orientation of the PCA relative to that magnetization, resulting in different polarizations of the THz emission from each source. Another possible way to control the THz emission could be to change the thickness of the materials or the separation between the two emitters to be on the order of a quarter wavelength. This would allow for controllable interference, at least at some frequencies. The wide range of parameters that can be tuned in such a hybrid emitter thus present significant opportunities for tailoring the THz pulse emitted.

\subsection{Emerging opportunities}
As the above section describes, coupling spintronic THz emitters to other materials platforms with complementary THz properties provides a large palette for designing multilayer structures whose THz performance exceeds that of any individual material. For example, Hu and coauthors predicted that THz radiation can be generated in transition metal dichalcogenides via second harmonic generation \cite{Hu2017}. Similarly, Welsh and coauthors have studied and observed the generation of terahertz radiation in metallic nanostructures like gold (Au) and silver (Ag) gratings, which relies on a resonant ``incoherent'' optical rectification \cite{welsh2009generation}. Those concepts could potentially also be employed in combination with spintronic heterostructures to further tailor the THz pulse characteristics. 

One class of material that is already being incorporated into multilayer THz heterostructures is topological insulators (TIs). TIs are materials that behave like insulators in the interior and conductors on the surface due to the presence of Dirac surface states that are topologically protected by time reversal symmetry\cite{zhu2015effect}. These surface states are known to have relativistic massless dispersion and have been probed using angle resolved photoemission spectroscopy (ARPES)\cite{zhu2015effect}. Terahertz generation from topological insulators largely originates in these Dirac surface states through two processes: the surface depletion field and optical rectification\cite{hsieh2009tunable, zhu2015effect}. Optical rectification is generally the dominant mechanism and results in a relatively large bandwidth due to optical transitions between the surface states and lower-energy conduction band states in the TI\cite{zhu2015effect}. For example, Zhu et al.~reported THz emission from pure Bi$_2$Se$_3$ thin film samples. The THz signal amplitude for samples with 10 quintuple layers was larger than that from samples with only 4 layers\cite{zhu2015effect}. This difference was attributed to strong coupling between the top surface and bottom surface of the TI in the thin samples. This coupling limited the surface-related optical transitions and suppressed the shift current\cite{zhu2015effect}. More recently, Wang et al.~reported THz emission with improved signal intensity in a Bi$_2$Se$_3$/Co heterostructure\cite{Wang_TI2018}. The improved performance was attributed to the strong spin–orbit coupling due to momentum-locked surface states leading to an enhanced ultrafast spin-to-charge current conversion in the heterostructure and thus enhanced THz emssion\cite{Wang_TI2018}.

\section{Conclusion}
Improved sources are crucial for advancing THz technology to take advantage of the many opportunities for sensing and spectroscopy in the THz frequency range. The purpose of this tutorial article was to introduce readers to spintronic THz emitters, which have emerged relatively recently as a powerful source of THz radiation with increased spectral bandwidth, intensity, and additional functionalities. Conceptually, the operation of these spintronic THz emitters is relatively simple. Because spintronic emitters take advantage of ultrafast dynamics in the spin degree of freedom to control the transient charge current, there are a wide variety of physical processes that can be leveraged to engineer these dynamics and the resulting subsequent THz emission. Our detailed description of the physical principles underlying the operation of the present generation of spintronic THz emitters is intended to help researchers new to the field to understand these opportunites. There is a wide range of material components, device geometries, and control strategies that can be explored and exploited to realize improved THz emission from spintronic and hybrid sources.

\section{Acknowledgement}
This research was primarily supported by NSF through the University of Delaware Materials Research Science and Engineering Center DMR-2011824. Additional support received from the NSF through Grant No. 1833000 and the University of Delaware Research Foundation.

\section*{Data Availability}
Data sharing is not applicable to this article as no new data were created or analyzed in this study.

\bibliographystyle{apsrev4-1}
\bibliography{ref}

\end{document}